\documentclass[journal=jctcce,manuscript=article]{achemso}
\setkeys{acs}{maxauthors=10,etalmode=truncate,chaptertitle=true,articletitle=true}

\usepackage[hidelinks]{hyperref}
\usepackage{amssymb}
\usepackage{amsmath}
\usepackage[dvipsnames]{xcolor}
\usepackage{graphicx}
\usepackage{float}
\usepackage[normalem]{ulem}
\usepackage{caption}
\usepackage{subcaption}
\usepackage{natbib}
\newcommand{\bra}[1]{\left\langle #1 \right|}
\newcommand{\ket}[1]{\left|#1\right\rangle}
\newcommand{\braket}[2]{\left\langle#1 |  #2\right\rangle}

\usepackage{xcolor}

\usepackage{multirow}
\usepackage{bm}
\usepackage[version=4]{mhchem} 
\usepackage{graphicx}
\usepackage{makecell}
\usepackage{booktabs}
\usepackage{enumitem}
\usepackage{siunitx}  
\newcolumntype{L}[1]{>{\raggedright\let\newline\\\arraybackslash\hspace{0pt}}m{#1}}
\newcolumntype{C}[1]{>{\centering\let\newline\\\arraybackslash\hspace{0pt}}m{#1}}
\newcolumntype{R}[1]{>{\raggedleft\let\newline\\\arraybackslash\hspace{0pt}}m{#1}}

\title{Extending orbital-optimized density functional theory to L-edge XPS and beyond:\\ Spin-orbit coupling via non-orthogonal quasi-degenerate perturbation theory}

\author{Richard Kang}
\email{richard.kang@berkeley.edu}
\altaffiliation{These authors contributed equally to this work.}
\affiliation{Kenneth S. Pitzer Center for Theoretical Chemistry, Department of Chemistry, University of California, Berkeley, CA 94720, USA}
\alsoaffiliation{Chemical Sciences Division, Lawrence Berkeley National Laboratory, Berkeley, CA 94720, USA}

\author{Leonardo A. Cunha}
\email{leonardo.cunha@berkeley.edu}
\altaffiliation{These authors contributed equally to this work.}
\affiliation{Kenneth S. Pitzer Center for Theoretical Chemistry, Department of Chemistry, University of California, Berkeley, CA 94720, USA}
\alsoaffiliation{Chemical Sciences Division, Lawrence Berkeley National Laboratory, Berkeley, CA 94720, USA}
\alsoaffiliation{Center for Computational Quantum Physics, Flatiron Institute, 162 5th Ave., New York, NY 10010, USA}

\author{Diptarka Hait}
\email{diptarka.hait@columbia.edu}
\affiliation{Department of Chemistry, Columbia University, New York, NY 10027, USA}
\alsoaffiliation{Initiative for Computational Catalysis, Flatiron Institute, New York, NY 10010, USA}
\alsoaffiliation{Department of Chemistry and The PULSE Institute, Stanford University, Stanford, CA 94305, USA}
\alsoaffiliation{SLAC National Accelerator Laboratory, Menlo Park, CA 94024, USA}
\author{Martin Head-Gordon}
\email{m_headgordon@berkeley.edu}
\affiliation{Kenneth S. Pitzer Center for Theoretical Chemistry, Department of Chemistry, University of California, Berkeley, CA 94720, USA}
\alsoaffiliation{Chemical Sciences Division, Lawrence Berkeley National Laboratory, Berkeley, CA 94720, USA}

\begin{document}
\maketitle
\begin{tocentry}
    \centering
\includegraphics[scale=1]{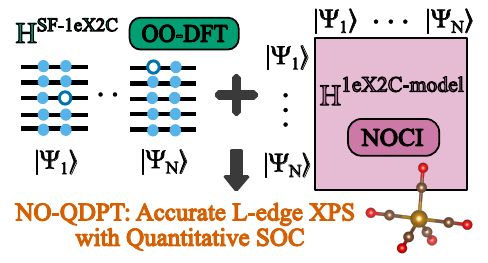} 
\end{tocentry}

\begin{abstract}
\noindent Quantum mechanical calculations of core electron binding energies (CEBEs) are relevant to interpreting x-ray photo-electron spectroscopy (XPS). 
Orbital-optimized density functional theory (OO-DFT) accurately predicts K-edge CEBEs but is challenged by the presence of significant spin-orbit coupling (SOC) at L- and higher edges involving inner-shell orbitals with nonzero angular momentum.
To extend OO-DFT to L-edges and higher, our method utilizes scalar-relativistic, spin-restricted open-shell OO-DFT to construct a minimal, quasi-degenerate basis of core-hole states corresponding to a chosen inner-shell (e.g. ionizing all six possible 2p spin orbitals).
Non-orthogonal configuration interaction (NOCI) is then used to obtain the matrix elements of the full Hamiltonian including SOC in this quasi-degenerate model space of determinants. Using a screened 1-electron SOC operator parametrized with the Dirac-Coulomb-Breit (DCB) Hamiltonian results in doublet splitting (DS) values for 3rd row elements that are nearly in quantitative agreement with experiment. The resulting NOCI eigenvalues are shifted by the average of the (scalar) OO-DFT CEBEs to yield CEBEs (split by SOC) corrected for dynamic correlation. Comparing calculations on gas phase molecules with experimental results establishes that NO-QDPT with the SCAN functional (NO-QDPT/SCAN), using the DCB screened 1-electron SOC operator is accurate to about 0.2 eV for L-edge CEBEs of molecules containing 3rd row atoms. 
However, this NO-QDPT approach becomes less accurate for 4th-row elements starting in the middle of the 3d transition metal series, with errors increasing as atomic number increases.

\end{abstract}

\section{Introduction}\label{sec:intro}

X-ray photoelectron spectroscopy (XPS) is a widely used analytical technique in the physical sciences.\cite{ratner2009electron, bagus2013interpretation} It dates back to the 1970s when Kai Siegbahn and Ulrik Gelius pioneered electron spectroscopy for chemical analysis (ESCA) which enabled the determination of core-electron binding energies of free molecules (and condensed matter).\cite{siegbahn1970angular, gelius1972esca,siegbahn1973esca,gelius1974recent, siegbahn1982electron, svensson1988electron} ESCA yielded crucial insights into core-ionization processes such as resonances, core-hole lifetimes, and gas-phase shake-up peaks, leading to Siegbahn sharing the 1981 Physics Nobel Prize ``for his contribution to the development of high-resolution electron spectroscopy." Today, XPS (the modern name for ESCA\cite{pinder2025s}) remains widely used for investigating condensed matter and interfacial phenomena.\cite{greczynski2020x} Some examples of its application include characterization of carbon-based materials\cite{okpalugo2005high, chen2020review}, energy storage devices, \cite{blyth2000xps, wood2018xps, shutthanandan2019applications} photoactive materials\cite{bhide1981depth, patrocinio2008xps} and bio-organic materials\cite{rouxhet2011xps}. 

The value of XPS stems from the element and shell specificity of the core-electron binding energies (CEBEs), which usually lie in a relatively narrow characteristic range. The precise location of the signals within this range reflects the local environment of each atom, such as oxidation state or the nature of chemical bonding.\cite{bagus2013interpretation,bagus2018extracting} Therefore reliable measurements of CEBEs provide key analytical information,\cite{greczynski2022step} whose interpretation can be greatly aided by  \textit{ab initio} CEBE calculations. We note that different sets of experimental CEBEs are subject to calibration errors\cite{crist2019xps, greczynski2020x} that have entered XPS electron binding energy data banks.\cite{crist2019xps}

The most widely used approaches for XPS calculations on molecules and condensed phase systems are computationally inexpensive ones based on density functional theory (DFT).\cite{kohn1965self,besley2021modeling,mardirossian2017thirty} The simplest approach is the direct use of ground-state orbital energies in a Koopman-like manner, but this typically yields very large errors with standard functionals.\cite{chong2002interpretation} Functionals with reduced self-interaction, however, lead to improved results.\cite{mei2018approximating,yu2025accurate} There are two commonly used alternatives. First is the well-established $\Delta$-SCF method\cite{bagus1965self,pedocchi1995co,cavigliasso1999accurate,vines2018prediction,kahk2019accurate,hait2020highly, hait2020accurate, carter2020state, garner2020core, zhao2021dynamic, kahk2021core} which captures key orbital relaxation effects (also referred to as orbital-optimized (OO)-DFT, particularly when generalized to excitations\cite{hait2021orbital}). With the aid of special solvers\cite{gilbert2008self, barca2018simple, hait2020excited,carter2020state} to converge to high-order saddle points in orbital space, OO-DFT methods predict highly accurate K-edge XPS (and X-ray absorption) spectra even for early transition metals,\cite{cunha2021relativistic} upon inclusion of scalar relativistic effects. Extensions of the OO approach to coupled cluster methods for K-shell CEBEs yield even higher accuracy.\cite{zheng2019performance,arias2022accurate} The second category of methods are based on Slater's transition state method,\cite{slater1972statistical} suitably generalized.\cite{williams1975generalization,chong1995accurate,triguero1999separate} Developments in this direction also continue.\cite{hirao2023core1s,jana2023slater,hirao2024verification}

In contrast to the K-edge features arising from ionization of 1s electrons, spin-orbit coupling (SOC) effects challenge calculations of core-ionized states from levels with non-zero angular momenta. 
For example, the 2p energy levels are split into $j=3/2$ and $j=1/2$ components by SOC according to how the unpaired electron spin $1/2$ couples to its orbital angular momentum of 1. Ionization from 2p levels therefore leads to two distinct peaks in the XPS spectrum, generally labeled as the lower energy L$_3$ (ionization out of $j=3/2$) and higher energy L$_2$ (ionization out of $j=1/2$) features. The separation of the two peaks is called doublet splitting (DS), which is another element-specific spectral signature from XPS. 

There have been relatively few reports of DFT calculations on 2p CEBEs.\cite{segala2006density,hait2020highly,besley2021density,hirao2023core2p} While one may use experimental DS values to model molecular L-edge spectra,\cite{hait2020highly, ross2024measurement, ou2024attosecond} \textit{ab initio} approaches that include SOC for L-edges are still an area of active research. Notably, nonrelativistic equation-of-motion coupled cluster single and doubles (EOM-CCSD) has been used to generate the set of relevant states, which were mixed under the Breit-Pauli SOC Hamiltonian to obtain L-edge spectra.\cite{epifanovsky2015spin, vidal2020equation} However, EOM-CCSD misses the orbital relaxation effects that make OO-DFT so successful for K-edges. On the other hand, non-orthogonal configuration interaction (NOCI)\cite{broer1981broken, sundstrom2014non, oosterbaan2018non, oosterbaan2019non, oosterbaan2020generalized} has been proposed as an alternative approach to couple relaxed orbital core-ionized configurations to compute L-edge XAS.\cite{grofe2022relativistic} While the main spectral features were present, the accuracy of that approach is limited by the lack of dynamic correlation.

Extending the success of OO-DFT for K-edges to L-edges and beyond is therefore an open challenge, with the key question being how to capture SOC, dynamic electron correlation and orbital relaxation together. In this work, we propose a solution towards that objective by combining OO-DFT and NOCI with a treatment of SOC via the exact 2-component (X2C)\cite{dyall1997interfacing, kutzelnigg2005quasirelativistic, saue2007inf, Liu2009x2c, cheng2011analytic} model. The resulting method has the advantage of addressing all three factors that are important for accurate L-edge calculations, as well as only requiring moderate computational effort. While our proposed approach is not fully \textit{ab initio}, highly encouraging results are obtained for calculations of L-edge energies and their DS values for molecules composed of main-group elements and transition-metal complexes. 
For example, the root mean squared error (RMSE) of DS and CEBE predictions are just 0.04 eV and 0.2 eV respectively for L-edges in molecules containing third-row main-group elements, vs experiment. 

The remainder of this paper is organized as follows. First we present an overview of the theory underlying our computational framework (Section \ref{sec:theory}), followed by a discussion of the main results in Section \ref{sec:results}, which include validation of the method for valence SOC splitting, and benchmarks across a range of molecules for 2p core-ionizations. We also explore the performance for extreme ultraviolet (XUV) accessible ionizations from M (3p and 3d) and N (4d) edges of a few heavier elements. We conclude in Section \ref{sec:conclusion} with a summary and future outlook.

\section{Theory and Implementation}
\label{sec:theory}

\subsection{One-Electron Exact Two Component Theory (1eX2C)}

One-electron exact two-component (1eX2C) theory decouples the large and small components of the four-component Dirac Hamiltonian.~\cite{dyall1997interfacing, kutzelnigg2005quasirelativistic, saue2007inf, Liu2009x2c} We direct interested readers to Refs \citenum{saue2011primer, Liu2012spin, dyallbook,cheng2011analytic} for further details and only provide a brief outline here. The 1eX2C approach is based on the one-electron operator $\hat{W} = (\hat{\vec{\sigma}} \cdot \hat{\vec{p}}) \hat{V}  (\hat{\vec{\sigma}} \cdot \hat{\vec{p}})$, where $\hat{p}$ is the electron momentum, $\hat{V}$ is the nuclear attraction, and $\hat{\vec{\sigma}}$ the vector of the Pauli operators. 
The use of a finite atom-centered basis reduces the single-particle Dirac equation in the restricted kinetic balance form\cite{kutzelnigg1984basis} to the following eigenproblem:
\begin{align}
\label{eq:dirac}
    \begin{bmatrix}
        \textbf{V} & \textbf{T} \\
        \textbf{T} & \dfrac{\textbf{W}}{4c^2} -\textbf{T} \\
    \end{bmatrix}
    \begin{bmatrix}
        \textbf{C}_L \\
        \textbf{C}_S \\
    \end{bmatrix} &= \epsilon
    \begin{bmatrix}
        \textbf{S} & 0 \\
        0 & \dfrac{1}{2c^2}\textbf{T} \\
    \end{bmatrix}
    \begin{bmatrix}
        \textbf{C}_L \\
        \textbf{C}_S \\
    \end{bmatrix} 
\end{align}
where $\textbf{T}$, $\textbf{V}$, and $\textbf{W}$ are the atomic orbital (AO) basis matrix representations of the operators for nonrelativistic kinetic energy ($\hat{T}$), $\hat{V}$, and $\hat{W}$. $\textbf{S}$ is the overlap matrix between the AO basis functions, $\textbf{C}_L$ is the large component and $\textbf{C}_S$ the small component. 

It is straightforward to decompose  $\hat{W}$ into spin-free $ \hat{W}^\textrm{SF}$ and spin-orbit $\hat{W}^\textrm{SO}$ parts, via: 
\begin{align}
    \hat{W} &= (\hat{\vec{\sigma}} \cdot \hat{\vec{p}}) \hat{V} (\hat{\vec{\sigma}} \cdot \hat{\vec{p}}) = \left( \hat{\vec{p}}\cdot \hat{V} \hat{\vec{p}} + i \hat{\vec{\sigma}} \cdot (\hat{\vec{p}} \times \hat{V} \hat{\vec{p}}) \right) \equiv \hat{W}^\textrm{SF} + i \hat{\vec{\sigma}} \cdot \hat{W}^\textrm{SO}  \label{eq:Wx2c2}
\end{align}
$\hat{W}^\textrm{SF}$ is spin-independent and leads to scalar relativistic effects.  On the other hand, the $i \vec{\sigma} \cdot \hat{W}^\textrm{SO}$ term is spin-dependent and does not commute with spin-operators. It thus couples different spin-states, leading to spin-orbit phenomena. 

Decoupling the large and small components of the Dirac Hamiltonian (Eq.~\ref{eq:dirac}) and writing it in a block-diagonal form requires defining the following transformations \cite{saue2011primer}: 
\begin{align}
\label{eq:decouple_x}
        \textbf{X} &= \textbf{C}_S (\textbf{C}_L)^{-1}\\
\label{eq:stilde}
    \tilde{\textbf{S}} &= \textbf{S} + \frac{1}{2c^2}\textbf{X}^\dagger \textbf{T} \textbf{X} \\
\label{eq:renorm_r}
    \textbf{R} &= \textbf{S}^{-1/2}\left( \textbf{S}^{-1/2} \tilde{\textbf{S}} \textbf{S}^{-1/2}\right)^{-1/2}\textbf{S}^{1/2}
\end{align}
which can be used to define the relativistic one-electron Hamiltonian matrix $\textbf{h}_1^{\rm X2C} = \textbf{T}^{\rm X2C} +  \textbf{V}^{\rm X2C}$ in terms of the effective kinetic energy $\textbf{T}^{\rm X2C}$ and nuclear attraction $ \textbf{V}^{\rm X2C}$ terms:
\begin{align}
\label{eq:t_x2c}
    \textbf{T}^{\rm X2C} &= \textbf{R}^\dagger\left(\textbf{T}\textbf{X} + \textbf{X}^\dagger \textbf{T} - \textbf{X}^\dagger \textbf{T} \textbf{X}\right) \textbf{R}\\
\label{eq:v_x2c}
    \textbf{V}^{\rm X2C} &= \textbf{R}^\dagger\left(\textbf{V} + \frac{1}{4c^2}\textbf{X}^\dagger \textbf{W} \textbf{X}\right) \textbf{R} 
\end{align}
We may also carry out the procedure described by Eqs.~\ref{eq:decouple_x}-\ref{eq:v_x2c} with only the spin-free part of $\hat{W}$ (\textit{i.e.,} explicitly replacing $\textbf{W}$ in  Eqn.~\ref{eq:dirac} and subsequent steps with $\textbf{W}^\textrm{SF}$), to define a scalar-relativistic one-electron Hamiltonian matrix $\textbf{h}_1^\textrm{SF-X2C} = \textbf{T}^\textrm{SF-X2C} +  \textbf{V}^\textrm{SF-X2C}$. 
We note that $\mathbf{T}^\textrm{SF-X2C}$ and $\mathbf{V}^\textrm{SF-X2C}$ differ from $\mathbf{T}^\textrm{X2C}$ and $\mathbf{V}^\textrm{X2C}$ as explicitly defined in Eqs.~\ref{eq:t_x2c}-\ref{eq:v_x2c}. This is due to the dependence of the transformation matrices $\textbf{X},\tilde{\textbf{S}},$ and $\textbf{R}$ on the nature of $\mathbf{W}$, i.e. whether only the spin-free term is used or the full operator.

With inclusion of the nonrelativistic electron-electron repulsion $\hat{G}^\textrm{NR}$ and the scalar nuclear-nuclear repulsion term $V_{nn}$, we can define the {full} 1eX2C and SF-1eX2C {Born-Oppenheimer molecular} Hamiltonians $\hat{H}^\textrm{1eX2C}$ and $\hat{H}^\textrm{SF-1eX2C}$ to be:
\begin{align}
\label{eqn:full1ex2c}
    \hat{H}^\text{1eX2C} &= \hat{h}_1^{\rm X2C}  + \hat{G}^{\rm NR} + V_{nn} = \hat{T}^{\rm X2C} + \hat{V}^{\rm X2C} + \hat{G}^{\rm NR} +V_{nn} \\
\label{eqn:sf1ex2c}
    \hat{H}^\textrm{SF-1eX2C} &= \hat{h}_1^\textrm{SF-X2C}  + \hat{G}^{\rm NR} + V_{nn} = \hat{T}^\textrm{SF-X2C} + \hat{V}^\textrm{SF-X2C} + \hat{G}^{\rm NR} +V_{nn} 
\end{align} 
The spin-orbit contribution in 1eX2C is therefore:
\begin{align}
\label{eqn:hx2csoc} 
    \hat{h}_1^\textrm{SOC-X2C} &= \hat{H}^\textrm{1eX2C}-\hat{H}^\textrm{SF-1eX2C}\equiv \hat{h}_1^\textrm{X2C}-\hat{h}_1^\textrm{SF-X2C}
\end{align}
without any two-electron contribution.
Exact two-component theory can also include two-electron terms, but their implementation is not straightforward (for discussion and alternatives, see Refs  \citenum{liu2018atomic, knecht2022exact, wang2025relativistic}). 
Using only spin-free one-electron terms also permits direct use of DFT exchange-correlation (XC) functionals~\cite{verma2016predicting, cunha2021relativistic}, 
as XC functionals do not depend on the one-electron Hamiltonian. However, the presence of spin-orbit terms requires (non-standard) non-collinear density functionals.{\cite{kubler1988density, sticht1989non, li2023noncollinear, pu2023noncollinear}} 

A simpler (but not \textit{ab initio}) approach to partially account for two-electron SOC effects stemmed from the observation that they screen the one-electron terms,\cite{botteger2000} which motivated empirical tuning of the one-body SOC contribution $\textbf{h}_1^\textrm{SOC-X2C}$. The Screened-Nuclear-Spin-Orbit (SNSO) approximation~\cite{botteger2000}, and subsequent modifications
~\cite{filatov2013spin, yoshizawa2016calculations} scale the matrix elements based on the angular momentum of the basis,
\begin{align}
    \textbf{h}_1^\textrm{SOC-SNSO} &\equiv \textbf{h}_1^\textrm{SOC-X2C} - \textbf{Q} \textbf{h}_1^\textrm{SOC-X2C} \textbf{Q} \\
    \textbf{h}_1^\textrm{X2C-SNSO} &\equiv \textbf{h}_1^\textrm{SF-X2C} + \textbf{h}_1^\textrm{SOC-SNSO} \label{eq:SNSO}
\end{align} 
where $\textbf{Q}$ is a diagonal matrix of angular-momentum-dependent coefficients. New SNSO coefficients based on four-component Dirac-Coulomb-Breit (DCB) Hamiltonian orbital splitting results are available in two forms\cite{ehrman2023improving}:
either universal (DCB-SNSO) or dependent on the row of the periodic table (rDCB-SNSO). In this work, we apply SNSO via Eq.~\ref{eq:SNSO}, although scaling $\textbf{W}_{SO}$  (and not $\textbf{h}_1^\textrm{SOC-X2C}$) has also been used {elsewhere in the literature}.~\cite{filatov2013spin, yoshizawa2016calculations, zou2020analytic} An alternative is to use empirical effective nuclear charge scaling, $Z_\text{eff}$, in the electron-nuclear attraction term.~\cite{koseki1992mcscf, koseki1995, koseki1998} 
We describe the use of $Z_\text{eff}$ charges to compute the one-body model X2C operator $\hat{h}_1^{\rm X2C-Z_{eff}}(Z_\text{eff})$ in Section S1 of the Supporting Information. Both the SNSO and the $Z_\text{eff}$ approaches lead to many-body X2C model relativistic Hamiltonians of the form: 
\begin{align}
\hat{h}_1^\textrm{X2C-model}&=\hat{h}_1^\textrm{SF-X2C}+ \hat{h}_1^\textrm{SOC-model}\\
\label{eqn:soc_all}
    \hat{H}^\textrm{1eX2C-model} &\equiv \hat{h}_1^\textrm{X2C-model} + \hat{G}^{\rm NR} +V_{nn} 
\end{align} 
A summary of key operators including the variants of one-electron Hamiltonians and the Born-Oppenheimer molecular Hamiltonians is provided in Table~\ref{tab:notation}. Further details on the two-electron SOC terms as well as their approximations 
can be found in Refs~\citenum{bernd1996, sikkema2009, liu2018atomic, knecht2022exact, wang2025relativistic}.

\renewcommand{\arraystretch}{1.15}
\begin{table}[!t]
\centering
\begin{tabular}{>{\(}l<{\)} >{\raggedright\arraybackslash}p{0.55\textwidth}}
\hline \hline
\textbf{Operator} & \textbf{Description} \\
\midrule
\multicolumn{2}{c}{\textbf{Fundamental Operators}} \\
\midrule
\hat{V} & Nonrelativistic nuclear attraction potential. \\
\hat{T} & Nonrelativistic kinetic energy. \\
\hat{G}^\textrm{NR} & Nonrelativistic electron-electron repulsion. \\
V_{nn} & Nonrelativistic nuclear-nuclear repulsion (scalar). \\
\hat{W}=(\hat{\vec{\sigma}} \cdot \hat{\vec{p}}) \hat{V}  (\hat{\vec{\sigma}} \cdot \hat{\vec{p}}) & One-electron operator central to 1eX2C. \\
\hat{W}^\textrm{SF}=\hat{\vec{p}}\cdot \hat{V} \hat{\vec{p}} & Scalar relativistic component of $\hat{W}$. \\
\hat{W}^\textrm{SO}=i \hat{\vec{\sigma}} \cdot (\hat{\vec{p}} \times \hat{V} \hat{\vec{p}})  & Spin-orbit component of $\hat{W}$. \\
\midrule
\multicolumn{2}{c}{\textbf{One electron Hamiltonians for X2C}} \\
\midrule
\hat{h}_1^{\rm X2C} & One-electron X2C Hamiltonian. \\
\hat{h}_1^\textrm{SF-X2C} & Scalar-relativistic one-electron X2C Hamiltonian. \\
\hat{h}_1^\textrm{SOC-X2C}=\hat{h}_1^{\rm X2C}-\hat{h}_1^\textrm{SF-X2C} & Spin-orbit contribution to the one-electron X2C Hamiltonian. \\
\hat{h}_1^\textrm{X2C-model}=\hat{h}_1^\textrm{SF-X2C}+ \hat{h}_1^\textrm{SOC-model}& One-electron X2C Hamiltonian including a model spin-orbit term. \\
\hat{h}_1^\textrm{X2C-SNSO}=\hat{h}_1^\textrm{SF-X2C}+ \hat{h}_1^\textrm{SOC-SNSO} & One-electron X2C Hamiltonian including a SNSO-screened SOC term. \\
\hat{h}_1^{\rm X2C-Z_{eff}}=\hat{h}_1^\textrm{SF-X2C}+ \hat{h}_1^{\rm SOC-Z_{eff}} & One-electron X2C Hamiltonian including a $Z_\text{eff}$-screened SOC term. \\
\midrule
\multicolumn{2}{c}{\textbf{Born-Oppenheimer molecular Hamiltonians}} \\
\midrule
\hat{H}^\textrm{1eX2C}= \hat{h}_1^{\rm X2C}  + \hat{G}^{\rm NR} + V_{nn} & Molecular Hamiltonian within the 1eX2C framework. \\
\hat{H}^\textrm{SF-1eX2C}= \hat{h}_1^\textrm{SF-X2C}  + \hat{G}^{\rm NR} + V_{nn} & Molecular Hamiltonian within SF-1eX2C. \\
\hat{H}^\textrm{1eX2C-model} = \hat{h}_1^\textrm{X2C-model}  + \hat{G}^{\rm NR} + V_{nn} & Molecular Hamiltonian for a spin-orbit model (e.g., SNSO or $Z_\text{eff}$) based on 1eX2C. \\
\hline \hline
\end{tabular}
\caption{Summary of key operators.}
\label{tab:notation}
\end{table}

\subsection{Non-Orthogonal Treatment of Spin-Orbit Coupling}\label{sec:noci-qdpt}

NOCI\cite{broer1981broken,thom2009hartree,sundstrom2014non} diagonalizes the Hamiltonian within a subspace that is typically spanned by a set of individually optimized determinants that are hence not necessarily orthogonal to each other. This permits greater flexibility than traditional CI based on a single set of orthogonal orbitals.
NOCI has been utilized to restore broken spatial symmetry\cite{broer1981broken,oosterbaan2018non} 
and to simulate core-hole states\cite{oosterbaan2018non,oosterbaan2019non, oosterbaan2020generalized,arias2024generalization,broer1981broken,grofe2022relativistic}. 

A fully variational treatment of SOC in our individual determinants is of course desired. However, $\hat{H}^\text{1eX2C-model}$ requires complex, spin-generalized orbitals. The {scarcity} of well-characterized non-collinear DFT functionals leaves Hartree-Fock (HF) as the {most widely applicable} spin-generalized mean-field approach available for orbital optimization with SOC.
We will sidestep this limitation by performing orbital optimization at the scalar relativistic level for 2p ionization from a given atom and then subsequently incorporate SOC through the NOCI Hamiltonian. This seems justified because SOC is quite local in space (decaying as $\sim O(r^{-3})$), and the SOC interaction strength is much smaller than typical energy differences between different orbital subshells of a given atom.
{For example, the 2p subshell of Ar is $\sim 250$ eV below the continuum (and the energetically closest subshell is Ar 2s at $\sim 330$ eV), while the atomic DS is $\sim 2$ eV.} We will therefore include SOC within the subspace of nearly degenerate states arising from local electron permutations within the relevant subshell: (\textit{e.g.,} the 6 core-hole states arising from {2p ionization on a given Ar atom}). We shall refer to this formalism as Non-Orthogonal Quasi-Degenerate Perturbation Theory (NO-QDPT).
We note that a similar rationale has been utilized in state-interaction based SOC treatments, such as in L-edge spectrum calculations in Ref. \citenum{lin2024elucidating} utilizing SF-1eX2C EOM-CCSD states\cite{cheng2018perturbative} and DFT/CIS L-edge calculations in Ref. \citenum{mandal2025computing} utilizing the Breit-Pauli Hamiltonian.

Given a ground state determinant $\ket{\Psi_{0}}$ and individually optimized core-hole determinants $\ket{\Psi_{1}}, \ket{\Psi_{2}}, \cdots \ket{\Psi_{N}}$, we compute the SOC effects through NOCI as follows: 
\begin{enumerate}
    \item Calculate overlap and Hamiltonian matrix elements within the core-hole state subspace: 
    \begin{align}
         {H}_{AB}^\textrm{1eX2C-model} &= \bra{\Psi_A} \hat{H}^\textrm{1eX2C-model} \ket{\Psi_B} &\text{ where } A,B \in \{1 \ldots N \} \\ 
        {S}_{AB} &= \braket{\Psi_A}{\Psi_B} &\text{ where } A,B \in \{1 \ldots N \}  .
    \end{align}
    \item Solve the generalized eigenvalue equation for core-hole states:
    \begin{align}
        \mathbf{H}^\textrm{1eX2C-model} \mathbf{C}^\textrm{1eX2C-model} = E^\textrm{1eX2C-model} \mathbf{S} \mathbf{C}^\textrm{1eX2C-model} 
        \label{eqn:geneval}
    \end{align} to obtain the energies $\{E_i^\text{1eX2C-model}\}$ and states $\left\{\displaystyle\sum\limits_A C_{i,A}^\textrm{1eX2C-model}\ket{\Psi_A}\right\}, i \in \{1..N\}$. This procedure not only accounts for SOC, but also any splitting of inner-shell orbitals arising from non-spherically symmetric environments in molecular systems\cite{JJohnson_1997,travnikova2006disentangling}.
    \item Utilize the ground state energy $E^\textrm{1eX2C-model}_0 = \bra{\Psi_{0}}\hat{H}^\textrm{1eX2C-model}\ket{\Psi_{0}}$ to compute observable energy differences like inner-shell ionization energies etc. 
\end{enumerate}
This procedure is agnostic about \textit{how} the determinants used in NOCI are obtained. The use of $\ket{\Psi_{0}},\ket{\Psi_{1}}, \ket{\Psi_{2}}, \cdots \ket{\Psi_{N}}$ optimized with SF-1eX2C using HF defines NO-QDPT/HF: a wavefunction theory that incorporates SOC for core-hole state related observables but lacks dynamic electron correlation. In principle, however, it is possible to use any well-defined set of determinants, as we discuss next.

\subsection{Shifting NOCI energies by OO-DFT\label{sec:dftshift}}

The formation of core-holes leads to a significant change in electron density, which can be effectively captured by direct optimization of orbitals for the core-hole state\cite{bagus1965self, banna1977study, besley2009self, kahk2019accurate, hait2020highly, cunha2021relativistic}. Such optimizations have been hindered in the past by the risk of ``variational collapse'' to lower energy solutions, as core-hole states are seldom (if ever) minima of energy in orbital space. Recent efforts to develop reliable algorithms for optimizing saddle point solutions in DFT \cite{gilbert2008self, barca2017excitation, hait2020excited, shea2020generalized, levi2020variational, carter2020state} have been key to wider use of OO-DFT, with the SCAN\cite{SCAN} functional typically having an RMSE of $\sim 0.3$ eV vs experimental energies in the soft X-ray regime\cite{hait2020highly, hait2020accurate, cunha2021relativistic}. 
The larger errors obtained from OO-HF as compared to OO-DFT (\textit{e.g.,} K-edge with SCAN\cite{hait2021orbital}) indicate that it would be desirable to incorporate some dynamic correlation effects from suitable density functionals into prediction of inner-shell properties. 

\begin{figure}[htb!]
    \centering
    \includegraphics[width=0.5\linewidth]{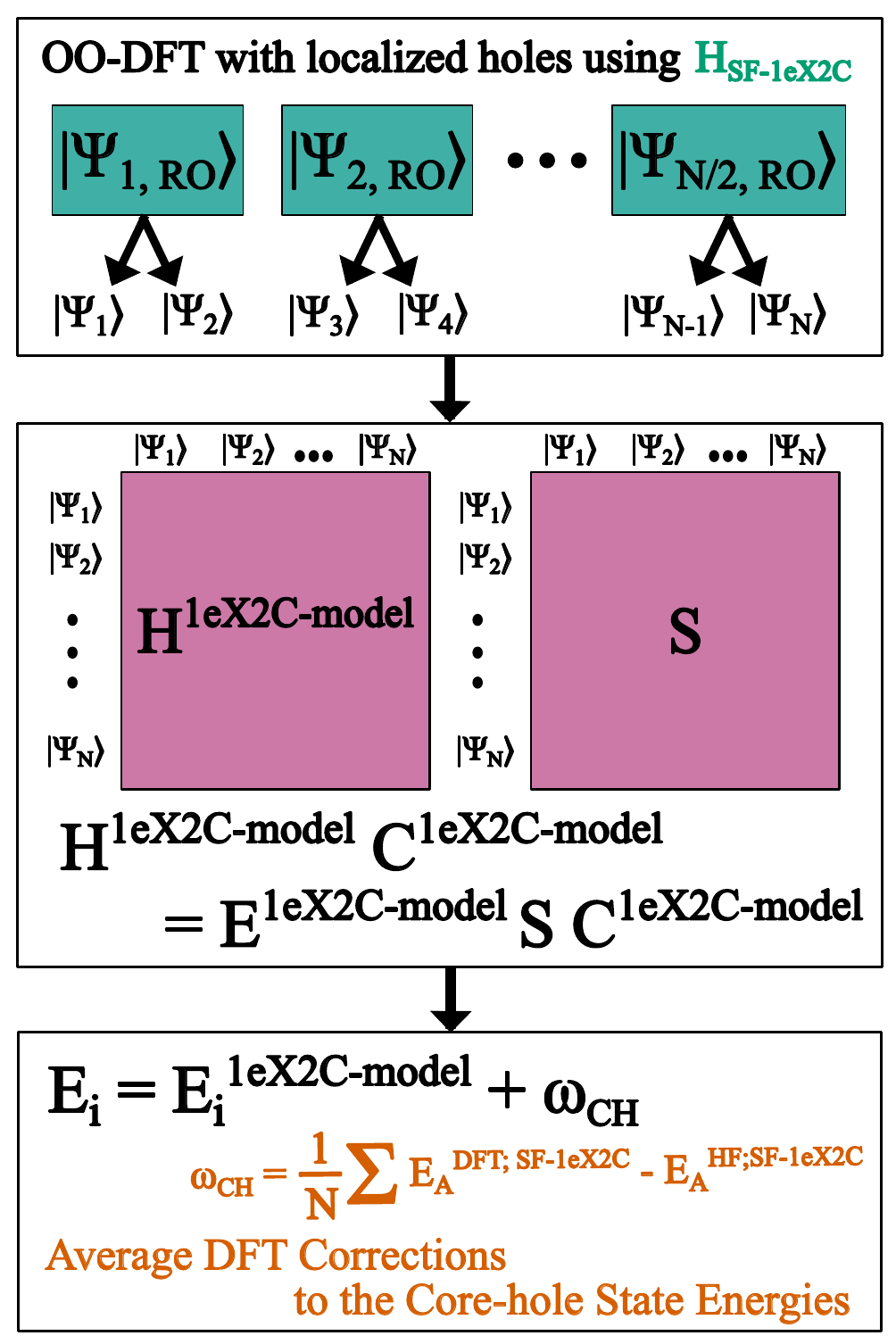}
\caption{Flowchart for NO-QDPT using both OO-DFT and NOCI. (top) $N/2$ restricted open-shell (RO) OO-DFT calculations are performed using $\hat{H}^\text{SF-1eX2C}$. Spin-permutation is used to obtain two states from one $\Delta \text{ROHF}$. (center) NOCI eigenvalue problem is solved in the basis of $N$ multi-electron states. (bottom) DFT corrections to the NOCI energies are applied to both ground state and core-hole states, although ground state correction is not described in the flowchart.} 
\label{fig:flowchart}
\end{figure}

We therefore propose the following NO-QDPT/DFT scheme (also described in Fig. \ref{fig:flowchart}). It is a hybrid of the pure wavefunction approach and OO-DFT in which the multiplet \textit{average} for a given subshell is obtained from OO-DFT and the subsequent spin-orbit splitting is obtained from wavefunction-based NOCI:
\begin{enumerate}
    \item Use DFT calculations to obtain the ground state determinant $\ket{\Psi_{0}}$ and core-hole determinants $\ket{\Psi_{1}}, \ket{\Psi_{2}}, \cdots \ket{\Psi_{N}}$, via orbital optimization to extremize the energy functional corresponding to $\hat{H}^\textrm{SF-1eX2C}$.  Let the resulting OO-DFT energies be $E_A^\text{DFT;SF-1eX2C}$ where $A\in \{0 \ldots N \}$.
    \item Evaluate the wavefunction theory (HF) energies of these determinants to obtain
    \begin{equation}
     E_A^\text{HF;SF-1eX2C} = \bra{\Psi_A} \hat{H}^\textrm{SF-1eX2C} \ket{\Psi_A}   
    \end{equation}
    Then evaluate the shift for the ground state ($\omega_0$) and the average shift for the core-hole state energies ($\omega_{\rm CH}$):
    \begin{align}
        \omega_0  & = E_0^\text{DFT;SF-1eX2C}-E_0^\text{HF;SF-1eX2C}\\
        \omega_{\rm CH} & =  \dfrac{1}{N}\displaystyle\sum\limits_{A>0} \left(E_A^\text{DFT;SF-1eX2C}-E_A^\text{HF;SF-1eX2C}\right) 
    \end{align}
    
    \item Calculate the overlap and Hamiltonian matrix elements within the core-hole state subspace from a wavefunction perspective: 
    \begin{align}
         {H}_{AB}^\text{1eX2C-model} &= \bra{\Psi_A} \hat{H}^\text{1eX2C-model} \ket{\Psi_B} &\text{ where } A,B \in \{1 \ldots N \} \\ 
        {S}_{AB} &= \braket{\Psi_A}{\Psi_B} &\text{ where } A,B \in \{1 \ldots N \}  .
    \end{align}
    \item Solve the generalized eigenvalue equation for core-hole wavefunctions:
    \begin{align}
        \mathbf{H}^\text{1eX2C-model} \mathbf{C}^\text{1eX2C-model} = E^\text{1eX2C-model} \mathbf{S} \mathbf{C}^\text{1eX2C-model} 
        \label{eqn:geneval_dft}
    \end{align} to obtain the NOCI energies $\{E_i^\text{1eX2C-model}\}$ and states $\left\{\displaystyle\sum\limits_A C_{i,A}^\textrm{1eX2C-model}\ket{\Psi_A}\right\}$
    \item Obtain the ground state SOC containing wavefunction energy via:
    \begin{align}
        E^\text{1eX2C-model}_0 = \bra{\Psi_{0}}\hat{H}^\text{1eX2C-model}\ket{\Psi_{0}}
    \end{align}
    \item Shift the SOC containing wavefunction energies via:
    \begin{align}
        E_0 & = E^\text{1eX2C-model}_0 +\omega_0 = E_0^\text{DFT;SF-1eX2C} + \bra{\Psi_0} \hat{h}_1^\text{SOC-model} \ket{\Psi_0}\\
        E_i & = E^\text{1eX2C-model}_i + \omega_{\rm CH} &\,  i \in \{1 \ldots N \}  
    \end{align}
    \item The core-ionization energies are thus $E_i-E_0$ etc. Note that the multiplet splittings $E_j-E_i= E^\text{1eX2C-model}_j- E^\text{1eX2C-model}_i$ ($i,j >0$) are not affected by $\omega_{CH}$ (and hence are independent of DFT energies). {Similarly, any splitting of inner-shell orbitals arising from a non-spherically symmetric molecular environment is unaffected by the average shift and is treated purely at the NOCI level. As a result, dynamic correlation effects are not accounted for in either the multiplet or molecular field splittings, but such contributions are expected to be quite small.\cite{Atanasov2012}}
\end{enumerate}

\subsection{Computational Details}

The protocols described above were implemented in a development version of the Q-Chem package\cite{epifanovsky2021software}, which was used for all calculations. We used square gradient minimization (SGM\cite{hait2020excited}) to converge restricted open-shell (RO) core-ionized, orbital-optimized determinants. Use of RO orbitals avoids potential spin-contamination problems that may have downstream impacts on SOC inclusive NOCI calculations. DFT calculations used a radial grid with 99 points and an angular Lebedev grid with 590 points. A total of 8 mean-field methods for Slater determinant optimization were tested: HF, SCAN\cite{SCAN}, SCAN hybridized with 10\% HF exchange (SCANh), SCAN0\cite{SCAN0}, B3LYP\cite{b3lyp}, $\omega$B97X-D3\cite{wB97XD3}, $\omega$B97X-V\cite{wb97xv}, and BHHLYP\cite{bhhlyp}. These functionals were chosen based on previous works which report the performance of OO-DFT for K-edges of main group elements.\cite{hait2020highly, hait2020accurate, hait2021orbital, cunha2021relativistic}

As mentioned above, exploiting the locality of core orbitals and SOC, we utilize the one-center approximation to NOCI\cite{oosterbaan2020generalized} for efficiency. The core-hole was thus localized onto a single atom for species with equivalent atoms (\textit{e.g.,} Cl in \ce{CCl4}), which prevents errors arising from delocalization\cite{perdew1982density, hait2018delocalization} of the hole over multiple sites\cite{hait2020highly}. We did not separately optimize the $\alpha$ and $\beta$ core-hole ionized states for closed-shell molecules -- we simply obtain one and flip the spin of the unpaired electron to generate the other.

The open-shell core-hole determinants are pre-processed before spanning the $\hat{H}^\textrm{1eX2C-model}$ space. SOC induces mixing between spin-pure states, and the $N_{\rm AO}\times N_{\rm AO}$ real-valued coefficient/density matrices from restricted open-shell optimizations have to be cast into $2N_{\rm AO}\times 2N_{\rm AO}$ sized matrices of complex numbers.  A finite difference approach~\cite{cunha2021relativistic} to computing the $\textbf{W}$ matrix elements was used for 1eX2C, for both the scalar-relativistic and spin-orbit parts. This scheme does not produce significant errors compared to analytical approaches.~\cite{cunha2021relativistic} More details on computational aspects of NO-QDPT are available in Section S2 of the Supporting Information.

Gas phase experimental geometries from NIST CCCBDB\cite{johnson2015nist} were used whenever possible. For molecules without experimental geometries, either MP2/cc-pVTZ structures from Ref \citenum{hait2020highly} were taken (\ce{P(CH3)4, Si(OCH3)4}), or new structures were optimized using $\omega$B97M-V\cite{wB97MV}/def2-TZVPPD\cite{weigend2005balanced}. All geometries and their sources are provided in the
Supporting Information in \texttt{XYZ} format.

\section{Results and Discussion}\label{sec:results}

For brevity, henceforth we denote the 1eX2C based Born-Oppenheimer molecular Hamiltonians with particular screening models as:
\begin{itemize}[itemsep=-0.5em]
    \item The unscreened 1eX2C Hamiltonian is denoted as X2C.
    \item The screened 1eX2C-DCB and 1eX2C-rDCB Hamiltonians are denoted as DCB-SNSO and rDCB-SNSO, respectively.
    \item The screened 1eX2C-$Z_\text{eff}$ Hamiltonian is denoted as $Z_\text{eff}$.
\end{itemize}

\subsection{Atomic Halogen Valence $^2P$ State Splittings\label{sec:pstates}}

\renewcommand{\arraystretch}{1.25}
\begin{table}[!htb]
\footnotesize
\begin{tabular}{l|l|R{0.5in}|R{0.5in}|R{0.5in}|R{0.5in}}
\hline \hline
\multicolumn{2}{c|}{} & \multicolumn{1}{c|}{F} & \multicolumn{1}{c|}{Cl} & \multicolumn{1}{c|}{Br} & \multicolumn{1}{c}{I} \\
\hline
\multicolumn{2}{c|}{\textbf{Expt. (cm$^\text{-1}$)}} & 404 & 881 & 3685 & 7602 \\
\hline \hline
\textbf{Method}  &  \multicolumn{1}{c|}{\textbf{Model} (${\hat{H}}$)} & \multicolumn{4}{c}{\textbf{Error (\%)}} \\
\hline
HF& X2C& 42.4& 13.3& 0.8& -4.5\\
SCAN& X2C& 47.3& 22.1& 11.1& 3.6\\
SCANh& X2C& 46.8& 21.4& 10.2& 3.1\\
SCAN0& X2C& 46.1& 20.2& 8.8& 2.3\\
B3LYP& X2C& 47.4& 23.9& 11.0& 3.7\\
$\omega$B97X-D3& X2C& 46.6& 23.2& 10.8& 4.1\\
$\omega$B97X-V& X2C& 45.8& 22.3& 10.8& 4.7\\
BHHLYP& X2C& 45.8& 20.4& 7.6& 1.2\\
\hline
HF& DCB-SNSO& -4.6& -6.5& -7.8& -9.8\\
SCAN& DCB-SNSO& -1.3& 0.8& 1.6& -2.2\\
SCANh& DCB-SNSO& -1.7& 0.2& 0.8& -2.7\\
SCAN0& DCB-SNSO& -2.1& -0.8& -0.4& -3.4\\
B3LYP& DCB-SNSO& -1.3& 2.3& 1.6& -2.1\\
$\omega$B97X-D3& DCB-SNSO& -1.8& 1.7& 1.4& -1.7\\
$\omega$B97X-V& DCB-SNSO& -2.3& 0.9& 1.4& -1.2\\
BHHLYP& DCB-SNSO& -2.3& -0.6& -1.5& -4.4\\
\hline \hline
\end{tabular}

\caption{Gas phase ground state halogen atom splittings calculated with {NO-QDPT}. Orbitals were computed using the decontracted aug-cc-p$\omega$CVTZ~\cite{dunning1989gaussian, kendall1992electron, woon1993gaussian, peterson2002accurate} (\ce{F, Cl}) and decontracted aug-cc-p$\omega$CVTZ-DK~\cite{wilson1999gaussian, de2001parallel, deyonker2007systematically, peterson2010molecular, bross2015correlation} (\ce{Br, I}). The experimental values are obtained from NIST Atomic Spectra Database~\cite{NISTASD}. The experimental data is in wavenumbers (cm\textsuperscript{-1}) and the calculated errors are reported in percent (\%). 
Experimental {uncertainties in} the reported values are 0.002 cm$^{-1}$ for F~\cite{laguna1982direct}, $<$0.1 cm$^{-1}$ for Cl~\cite{radziemski1969wavelengths}, 0.6655 cm$^{-1}$ for Br~\cite{tech1963analysis, davies1972gas} and 0.673 cm$^{-1}$ for I~\cite{kiess1959description, cerny1991experimental}. 
}
\label{tab:halogen}
\end{table}

To validate our approach and its implementation, we examined the valence splittings of atomic halogens. The ground state valence electronic configuration of the halogen atoms is ns$^2$np$^5$ ($^2P$). SOC splits this sixfold degeneracy into $^2P_{3/2}$ and $^2P_{1/2}$ levels. Highly accurate experimental DS values are available, ranging from 404 cm$^{-1}$ in F to 7603 cm$^{-1}$ in I. 
Table~\ref{tab:halogen} summarizes the errors (\%) in calculated {NO-QDPT} DS, combined with X2C and the universal DCB-SNSO SOC parametrization.\cite{ehrman2023improving}. The lowest errors for each halogen atom are obtained from SCAN, SCANh, SCAN0, and $\omega$B97X-V with DCB-SNSO, respectively, while unscreened SCAN/X2C leads to much larger errors. 
Results with the row dependent rDCB parameterization (rDCB) are similar to DCB (Supporting Information, Table S4-I).
The value of the DFT-based NO-QDPT approach for the splittings is made clear by comparing DFT/DCB-SNSO errors against HF/DCB-SNSO errors. Comparing SCAN and HF, for instance, the HF errors are larger by factors of 3.5 (F) to 8 (Cl). In fact for all halogen atoms, the HF/DCB-SNSO errors are at least double that of the worst performing DFT functional.
We reiterate that the NO-QDPT splittings arise purely through NOCI in the space of optimized determinants and are independent of DFT \textit{energies}. Therefore, the improvement from HF to SCAN is solely the result of improved DFT \textit{orbitals}. 
DFT orbitals also improve over HF in NO-QDPT predictions of alkali metal atom D-line splittings (Supporting Information, Section S4 and Table S4-I).   

\subsection{L$_\textrm{2}$ and L$_\textrm{3}$ Edges for 3\textsuperscript{rd} Row Elements \label{sec:3rdrow}}

\renewcommand{\arraystretch}{1.25}
\begin{table}[!t]
\footnotesize
\begin{tabular}{l|l|R{0.5in}|R{0.5in}|R{0.5in}|R{0.5in}|R{0.5in}}
\hline \hline
\multicolumn{2}{c|}{} & \multicolumn{1}{c|}{Ar} & \multicolumn{1}{c|}{HCl} & \multicolumn{1}{c|}{\ce{H2S}} & \multicolumn{1}{c|}{\ce{PH3}} & \multicolumn{1}{c}{\ce{SiH4}}\\
\hline
\multicolumn{2}{c|}{\textbf{Expt. DS (eV)}}  & 2.15\cite{king1977investigation} & 1.63\cite{shaw1984inner} & 1.20\cite{hudson1994high} & 0.90\cite{liu1990high} & 0.60\cite{hayes1972absorption} \\
\hline \hline 
\textbf{Method}  &  \multicolumn{1}{c|}{\textbf{Model} (${\hat{H}}$)} & \multicolumn{5}{c}{\textbf{Error (eV)}} \\
\hline 
Koopman cGHF& X2C& 0.49& 0.4& 0.33& 0.23& 0.21\\
$\Delta$SCF cGHF& X2C& 0.43& 0.35& 0.29& 0.19& 0.19\\
NO-QDPT HF& X2C& 0.37& 0.30& 0.25& 0.17& 0.16\\
NO-QDPT SCAN& X2C& 0.42& 0.34& 0.28& 0.19& 0.18\\
NO-QDPT SCAN0& X2C& 0.41& 0.33& 0.27& 0.18& 0.18\\
\hline 
Koopman cGHF& DCB-SNSO& 0.15& 0.13& 0.11& 0.06& 0.07\\
$\Delta$SCF cGHF& DCB-SNSO& 0.10& 0.08& 0.08& 0.02& 0.05\\
NO-QDPT HF& DCB-SNSO& -0.04& -0.03& -0.02& -0.04& 0.00\\
NO-QDPT SCAN& DCB-SNSO& -0.01& -0.01& 0.01& -0.03& 0.01\\
NO-QDPT SCAN0 & DCB-SNSO& -0.01& -0.01& 0.00& -0.03& 0.01\\
\hline \hline 
\end{tabular}
\caption{Gas phase 3\textsuperscript{rd} row element L$_\textrm{2,3}$-edge DS values for 18-electron series. All methods were computed using aug-pcX-2 basis set~\cite{ambroise2018probing} on 3\textsuperscript{rd} row elements and decontracted aug-pcseg-1~\cite{jensen2014unifying} for \ce{H}. Full results with the absolute L$_\textrm{2,3}$ IPs as well as the basis set screening results are available in Supporting Information, Table S4-II, III. The experimental DS uncertainties provided for each value are 0.05 eV for \ce{Ar}, 0.03 eV for \ce{HCl}, 0.01 eV for \ce{H2S}, 0.05 eV for \ce{SiH4}.
}
\label{tab:18eseries}
\end{table}

We now explore the applicability of our NO-QDPT approach for computing the 2p CEBEs of 3\textsuperscript{rd} row elements in gas phase molecules. 
NO-QDPT assumes that orbital relaxation in the presence of SOC is small. The validity of this assumption is tested via comparison to full $\Delta$cGHF (\textit{i.e.,} separate orbital optimization of 2p$_{3/2}$ and 2p$_{1/2}$ hole states with HF, in the presence of SOC) for four simple hydrides and the Ar atom (Table~\ref{tab:18eseries}). The computational details for converging $\Delta$cGHF calculations are provided in Supporting Information Section S3. The Koopman's orbital energy model for the DS (\textit{i.e.,}  $\varepsilon(\textrm{2p}_{3/2}) -\varepsilon(\textrm{2p}_{1/2})$ for the neutral cGHF ground state) is also considered, as is NO-QDPT/SCAN. Similar to the behavior observed for $^2P$ halogen atom valence splittings, Table~\ref{tab:18eseries} shows that X2C overestimates the L$_\textrm{2,3}$-edge DS values for all methods, with the errors growing with atomic number. By contrast, DCB-SNSO leads to much smaller errors ($<$ 0.1 eV).
The DCB-SNSO data for NO-QDPT/HF underestimates the DS by a nearly negligible 0.02-0.05 eV versus $\Delta$cGHF, which validates its design. 
Interestingly, NO-QDPT/SCAN leads to a slightly lower difference with full $\Delta$cGHF.
Somewhat surprisingly, the Koopman's model reproduces the $\Delta$cGHF DS very well, overestimating by $\lesssim 0.05$ eV. 
This further bolsters the claim that core-hole relaxation effects have a rather small effect on the DS for 3rd row elements. As expected, the absolute Koopman's L$_3$ and L$_2$ CEBEs are greatly overestimated ($\sim$10 eV errors) due to lack of core-hole relaxation (Supporting Information, Table S4-III).

\begin{figure}[t!]
    \centering
    \includegraphics[width=0.5\linewidth, trim=50 0 50 30]{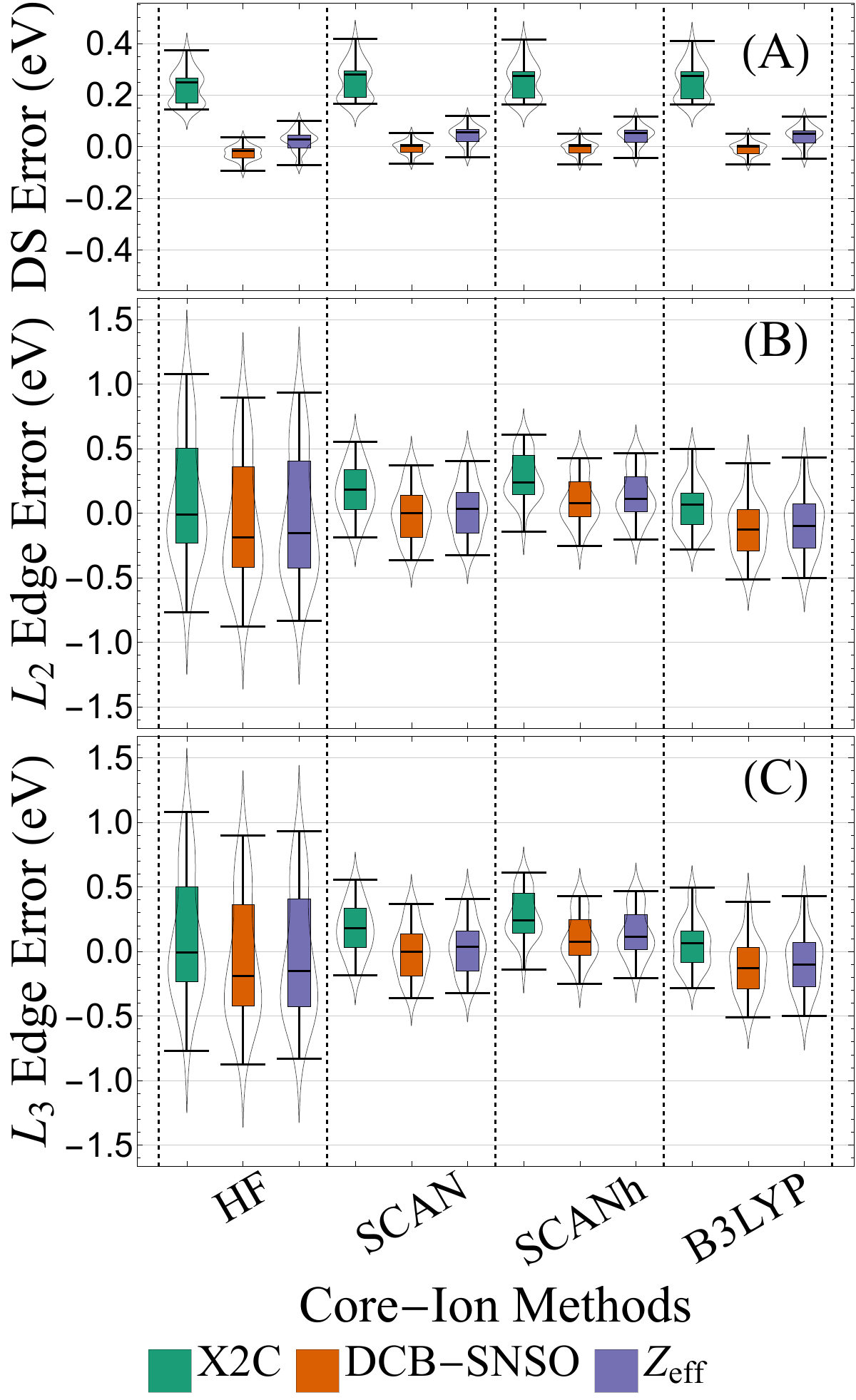}
\caption{Violin plots for the gas phase 3\textsuperscript{rd} row element L$_\textrm{2,3}$-edges. Boxes span the 50\% as the interquartile range with the median noted, and the underlying contours show the distribution of the data. All methods were computed using aug-pcX-2~\cite{ambroise2018probing} except for \ce{H} which used decontracted aug-pcSeg-1~\cite{jensen2014unifying} basis set. Individual data points, error, and RMSE for 25 molecules and 8 functionals are available in the Supporting Information along with their experimental references (Section S5, Fig S1 and Table S4-IV). SCANh stands for SCAN with 10\% exact exchange.}
\label{fig:3rdrow}
\end{figure}

We next consider L-edge ionization energies and DS for 25 closed-shell, gas-phase molecules containing 3rd row elements. Violin plots of the error distributions for L$_{2,3}$ ionization energies and the DS (compared to experiment) are reported in Fig.~\ref{fig:3rdrow}. In this application, the choice of optimized determinants (HF or DFT) has little impact on the DS errors. Again, unscreened X2C systematically overestimates the DS, while the screened models (DCB-SNSO and $Z_\text{eff}$) lead to much smaller errors (RMSE $\lesssim 0.06$ eV), with DCB performing slightly better than $Z_\text{eff}$. The largest DS overestimation is  for Ar (highest $Z$). The largest underestimation is for \ce{CCl4} (experimental DS\cite{aitken1980electron} of 1.69 eV). We note that another experiment\cite{hitchcock1978inner} reported a DS of 1.6 eV for this molecule, which is closer to the screened SOC results (as well as to the 1.6 eV separation between multiplet peaks observed in the L-edge X-ray absorption spectrum\cite{ross2024measurement,lu2008state}). 
This example aside, we do not observe any significant shifts in the DS (relative to the values for the simple hydrides in Table~\ref{tab:18eseries}) for a specific element due to the molecular environment.

In contrast to the behavior for predicting the DS, the error distributions for the L$_{2,3}$ CEBEs reveal the advantages of using DFT over HF, with most of the tested functionals (except BHHLYP) roughly halving the spread of the error distribution. Indeed, NO-QDPT/SCAN/DCB-SNSO yields an RMS error of 0.2 eV, which is comparable to results obtained for K-edges of similar molecules\cite{hait2020highly, cunha2021relativistic} and not much larger than typical experimental uncertainties ($\sim 0.1$ eV). B3LYP with DCB-SNSO also yields a similar performance with RMS error of 0.2 eV. These are very encouraging results. By contrast, NO-QDPT/HF/DCB-SNSO has a larger RMS error of $0.5$ eV, due to lack of dynamic electron correlation. The choice of the functional matters for NO-QDPT/DFT: BHHLYP yields worse performance than the SCAN family of functionals and SCAN0 slightly degrades the good performance of SCAN. In addition, while BHHLYP mostly overestimates CEBEs, the two tested range-separated hybrids ($\omega$B97X-D3 and $\omega$B97X-V) tend to underestimate the CEBEs. The full results for the tested functionals not shown in Fig.~\ref{fig:3rdrow} is given shown in  Supporting Information, Section S5 Figure S1.
We also stress that the success of NO-QDPT/DFT depends on more than choosing a suitable functional. It also relies on the validity of the relativistic $\hat{H}^\textrm{model}$ given by Eq. \ref{eqn:soc_all}, and the quality of the NOCI subspace. For the latter, it is important to avoid linear dependencies within the NOCI subspace (some strategies for achieving this are provided in Supporting Information, Section S2).

\subsection{L$-$ Edges for 4th Row Main Group Elements \label{sec:4throw}}

\renewcommand{\arraystretch}{1.25}
\begin{table}[!t]
\footnotesize
\begin{tabular}{l|l|r|r|r|r|r|r|r}
\hline \hline
\multicolumn{2}{c|}{} & \multicolumn{1}{c|}{KCl} & \multicolumn{1}{c|}{KBr} & \multicolumn{1}{c|}{Ca} & \multicolumn{1}{c|}{Zn} & \multicolumn{1}{c|}{\ce{GeH4}} & \multicolumn{1}{c|}{\ce{GeCl4}} & \multicolumn{1}{c}{Kr} \\
\hline
\multicolumn{2}{c|}{\textbf{Expt. L$_3$ (eV)}} & 300.8\cite{patanen2012direct}& 301.1\cite{patanen2008strong}& 357.6\cite{wernet1998determination}& 1029.1\cite{banna1978free_zncd}& 1225.7\cite{venezia1987near}& 1228.1\cite{venezia1987near}& 1679.2\cite{dragoun2004increased} \\
 \hline \hline
\textbf{Method} & \multicolumn{1}{c|}{\textbf{Model} (${\hat{H}}$)}  & \multicolumn{7}{c}{\textbf{Error (eV)}}\\
\hline
HF& X2C& -0.1& -0.3& -0.6& -0.6& -0.3& 0.7& 0.3 \\
SCAN& X2C& -0.5& -0.8& -0.7& -2.8& -4.6& -4.2& -6.3\\
SCAN0  & X2C& -0.1& -0.3& -0.4& -2.0& -3.3& -2.8& -4.4\\
B3LYP  & X2C& -1.0& -1.2& -1.6& -4.0& -5.6& -6.9& -7.6\\
\hline
HF& DCB-SNSO& 0.1& -0.1& -0.4& 0.2& 0.7& 1.8& 1.8 \\
SCAN& DCB-SNSO& -0.3& -0.6& -0.5& -1.9& -3.5& -3.2& -4.7\\
SCAN0  & DCB-SNSO& 0.1& -0.1& -0.2& -1.2& -2.3& -1.7& -2.9\\
B3LYP  & DCB-SNSO& -0.8& -1.0& -1.4& -3.1& -4.6& -5.8& -6.1\\
\hline \hline
\multicolumn{9}{c}{} \\
\hline \hline
\multicolumn{2}{c|}{} & \multicolumn{1}{c|}{KCl} & \multicolumn{1}{c|}{KBr} & \multicolumn{1}{c|}{Ca} & \multicolumn{1}{c|}{Zn} & \multicolumn{1}{c|}{\ce{GeH4}} & \multicolumn{1}{c|}{\ce{GeCl4}} & \multicolumn{1}{c}{Kr} \\
\hline
\multicolumn{2}{c|}{\textbf{Expt. DS (eV)}} & 2.8\cite{patanen2012direct}& 2.8\cite{patanen2008strong}& 3.4\cite{wernet1998determination}& 23.2\cite{banna1978free_zncd}& 31.1\cite{venezia1987near} & 30.9\cite{venezia1987near}& 52.7\cite{dragoun2004increased} \\
 \hline \hline
\textbf{Method} & \multicolumn{1}{c|}{\textbf{Model} (${\hat{H}}$)} & \multicolumn{7}{c}{\textbf{Error (eV)}}\\
\hline
HF     & X2C& 0.5& 0.5& 0.7& 2.1& 2.4& 2.6& 3.2\\
SCAN   & X2C& 0.5& 0.5& 0.8& 2.4& 2.8& 3.0& 3.7\\
SCAN0  & X2C& 0.5& 0.5& 0.8& 2.3& 2.7& 2.9& 3.6\\
B3LYP  & X2C& 0.5& 0.5& 0.8& 2.4& 2.7& 2.9& 3.7\\
\hline
HF     & DCB-SNSO& 0.0& 0.0& 0.1& -0.4& -0.7& -0.5& -1.4\\
SCAN   & DCB-SNSO& 0.0& 0.0& 0.2& -0.2& -0.4& -0.2& -0.9\\
SCAN0  & DCB-SNSO& 0.0& 0.0& 0.1& -0.2& -0.5& -0.3& -1.0\\
B3LYP  & DCB-SNSO& 0.0& 0.0& 0.1& -0.2& -0.4& -0.2& -0.9\\
\hline \hline
\end{tabular}

\caption{Gas phase 4\textsuperscript{th} row element L\textsubscript{2,3}-edge DS and absolute L\textsubscript{3} edges computed through {NO-QDPT}. 
All methods were computed using decontracted cc-pVDZ for H and Cl, decontracted aug-cc-p$\omega$CVTZ-X2C\cite{hill2017gaussian} for K, Ca, and decontracted aug-cc-p$\omega$CVTZ-DK for Zn, Ge, Br, Kr.\cite{de2001parallel, deyonker2007systematically}
Full results with the absolute L$_\textrm{2,3}$ IPs are provided in Supporting Information, Table S4-VI.}
\label{tab:ledge_4throw}
\end{table}

We next consider 2p ionization behavior for molecules containing 4th row main group elements  (Table \ref{tab:ledge_4throw}). Since gas phase experimental data is scarce, we can report results for only a few species. We restricted our testing to HF, SCAN, SCAN0 and B3LYP, based on our results for the 3rd row main group elements.
The s block elements (K and Ca) behave similarly to the 3rd period elements in that unscreened X2C overestimates the DS, while DCB-SNSO is essentially quantitative. All of the tested methods yield quite accurate L$_{2,3}$ DS for K and Ca ($\sim$ 0.2 eV error). 
Notably, B3LYP shows slightly degraded performance in predicting L$_3$ ionization energies, compared to the other tested methods.

The heavier elements Zn, Ge, and Kr present a somewhat less encouraging picture, at first glance. For the DS, errors from the NO-QDPT/SCAN/DCB-SNSO model increase to as much as 1 eV for Kr.  
rDCB-SNSO Hamiltonian slightly degrades performance relative to DCB (Supporting Information, Table S4-VI), while Z$_\text{eff}$ is significantly poorer (Supporting Information Section S6).
While the absolute DS errors are considerably larger for these heavier elements, so too is the magnitude of the DS themselves. In percentage terms, the performance of NO-QDPT/DFT/DCB-SNSO is not significantly degraded, but the systematic underestimation hints at limitations of the hybrid protocol, the relativistic Hamiltonian, and/or the density functional. {In this regard, we note that $\Delta$cGHF calculations for the L$_{2,3}$ ionization energies of Kr find $<0.1$ eV error in the DS, compared to -1.5 eV from the NO-QDPT protocol with HF/DCB-SNSO and 0.4 eV from a Koopman energy difference of neutral Kr cGHF 2p orbitals. This indicates that the interplay between SOC and orbital relaxation is potentially non-negligible for these heavier elements, and a fully variational treatment may be optimal.} 

The performance for L$_{2,3}$ CEBEs for molecules containing Zn, Ge, and Kr is much worse than for K and Ca.  
All NO-QDPT/DFT/DCB-SNSO systematically underestimates the L$_{3}$ CEBEs by as much as 6.1 eV for Kr with B3LYP. SCAN0 underestimates the least of the three, but still yields considerable errors. Surprisingly, HF is the best-performing method, which is presumably fortuitous, and reflects other errors already mentioned. We note that in percentage terms, the worst NO-QDPT/DFT/DCB-SNSO error is less than 0.4\% for Kr (B3LYP). More importantly, the NO-QDPT/DFT/DCB-SNSO error in the \textit{shift} between \ce{GeH4} and \ce{GeCl4} L$_{3}$ CEBEs is only 0.2 eV out of 2.4 eV, so the theory still appears promising for chemical shift effects, despite the noticeable degradation in performance in predicting absolute CEBEs for these heavier elements.

\subsection{L$-$ Edges for 1\textsuperscript{st} Row Transition Metals \label{sec:tms}}

\renewcommand{\arraystretch}{1.25}
\begin{table}[!t]
\footnotesize
\begin{tabular}{l|c|r||R{0.4in}|R{0.4in}|R{0.4in}|R{0.4in}|R{0.4in}|R{0.4in}}
\hline \hline
\multicolumn{3}{c||}{}
 & \multicolumn{1}{c|}{HF} 
 & \multicolumn{1}{c|}{SCAN}
 & \multicolumn{1}{c|}{B3LYP} 
 & \multicolumn{1}{c|}{HF} 
 & \multicolumn{1}{c|}{SCAN}
 & \multicolumn{1}{c}{B3LYP} \\ 
 \hline
\multicolumn{2}{c|}{\textbf{Expt. L$_3$ (eV)}} & \multicolumn{1}{c||}{\textbf{OS}} 
   & \multicolumn{3}{c|}{\textbf{X2C Error (eV)}} & \multicolumn{3}{c}{\textbf{DCB-SNSO Error (eV)}} \\
\hline \hline
\ce{V(CO)4(C5H5)} & 519.6 & +1  & 1.3& -0.6& -1.3& 1.6& -0.3& -1.0\\
\ce{Cr(CO)6}& 582.0 & 0  & 0.3& -0.6& -1.7& 0.7& -0.2& -1.3\\
\ce{Cr(CO)5(CS)}& 581.7 & 0 & 0.5& -0.5& -1.6& 0.9& -0.1& -1.2\\
\ce{Mn(CO)5H}& 647.5 & +1  & 0.3& -0.8& -2.0& 0.8& -0.4& -1.6\\
\ce{Mn(CO)3(C5H5)}& 646.7 & +1  & 0.5& -0.9& -2.1& 1.0& -0.5& -1.7\\
\ce{Mn(CO)2(CS)(C5H5)} & 646.8 & +1  & 0.4& -1.0& -2.2& 0.8& -0.6& -1.8\\
\ce{Fe(CO)5}& 715.8 & 0  & -0.3& -0.8& -2.3& 0.2& -0.3& -1.8\\
\ce{Fe(CO)4(C2H4)}& 715.4 & 0  & 0.0& -1.1& -2.5& 0.5& -0.6& -2.0\\
\ce{Fe(CO)4(H)2}& 716.0 & +2 & 1.2& -0.8& -2.2& 1.7& -0.3& -1.7\\
\ce{Fe(C5H5)2}& 713.1 & +2 & 2.5& -0.9& -2.1& 3.0& -0.4& -1.6\\
\ce{Fe(H3CC5H5)2}& 713.0 & +2 & 2.3& -1.1& -2.3& 2.8& -0.6& -1.8\\
\ce{Co(CO)3(NO)}& 786.9 & -1 & 0.3& -1.0& -2.6& 0.8& -0.4& -2.0\\
\ce{Co(CO)4(H)}& 786.9 & +1 & 0.1& -0.8& 1.5& 0.7& -0.2& 1.5\\
\ce{Co(CO)2(C5H5)}& 786.3 & +1 & -0.3& -1.6& -3.1& 0.3& -1.0& -2.5\\
\ce{Ni(CO)4}& 861.2 & 0 & -2.5& -1.4& -3.3& -1.9& -0.7& -2.6\\
\ce{Ni(PF3)4}& 862.0 & 0 & -2.1& -2.0& -3.6& -1.5& -1.3& -3.0\\
\ce{Ni(acac)2}& 860.5 & +2 & 3.3 & -2.3 & -3.2 & 4.0 & -1.6 & -2.5 \\
\hline \hline
\end{tabular}
\caption{Transition Metal (TM) L$_3$ ionization energies for complexes that satisfy the 18-electron rule, calculated through {NO-QDPT}. The transition metal oxidation states (TM OS) are also provided. All methods were computed using decontracted aug-cc-p$\omega$CVTZ\cite{balabanov2005systematically} on TM centers and decontracted aug-cc-pVDZ basis set on others. The acetylacetonate ligand is abbreviated as acac. Full results with absolute L$_\text{2,3}$ IPs are available in Supporting Information, Table S4-VII.}
\label{tab:tml3edge_18e}
\end{table}

The differing quality of results for K/Ca versus Zn/Ge/Kr makes assessment of the performance of NO-QDPT/DFT for the first row transition metals very interesting. We first examine L$_\text{3}$ CEBEs where there is more available experimental data, followed by DS results on a few species with experimental values. 
Tables \ref{tab:tml3edge_18e} and \ref{tab:tml3edge_non18e} present the L$_3$ ionization energies in 18-electron and non-18-electron transition metal complexes, respectively. 

We first discuss the case of the 18-electron complexes (which presumably have less challenging ground state electronic structure). From Table \ref{tab:tml3edge_18e}, we see performance that roughly mirrors what was reported for the 4th row main group elements above. Notably NO-QDPT/SCAN/DCB-SNSO outperforms the corresponding HF approach, though both underestimate the CEBEs. As already seen above, the underestimation grows towards $>1 \text{eV}$ for TMs with increasing atomic numbers, such as Co and Ni, with percentage errors also increasing (albeit more slowly). Notably the accuracy of B3LYP is worse than SCAN both with or without screening of SOC. The situation is better for the relative chemical shifts than absolute CEBEs, with errors roughly halved. It is interesting to note that the errors in relative chemical shifts for a given element are little changed by replacing SCAN by HF, indicating that effects other than the quality of orbitals are responsible for these errors.

\renewcommand{\arraystretch}{1.25}
\begin{table}[!htb]
\footnotesize
\begin{tabular}{l|c|c|c||R{0.4in}|R{0.4in}|R{0.4in}|R{0.4in}|R{0.4in}|R{0.4in}}
\hline \hline
\multicolumn{4}{c||}{}
 & \multicolumn{1}{c|}{HF} 
 & \multicolumn{1}{c|}{SCAN}
 & \multicolumn{1}{c|}{B3LYP} 
 & \multicolumn{1}{c|}{HF} 
 & \multicolumn{1}{c|}{SCAN}
 & \multicolumn{1}{c}{B3LYP} \\ 
 \hline
\multicolumn{2}{c|}{\textbf{L$_3$ Expt. (eV)}} & \multicolumn{1}{c|}{\textbf{$e^{-}$ Count}} & \multicolumn{1}{c||}{\textbf{OS}} & \multicolumn{3}{c|}{\textbf{X2C Error (eV)}} & \multicolumn{3}{c}{\textbf{DCB-SNSO Error (eV)}} \\
\hline \hline
\ce{TiCl4}     & 465.4\cite{10.1063/1.436571} & 8 & +4  & 6.1 & -0.9 & -1.0 & 6.4 & -0.6 & -0.7 \\
\ce{Ti(NO3)4}  & 466.8\cite{FORD198375} & 8 & +4  & 6.7 & -1.9 & -1.5 & 7.0 & -1.6 & -1.2 \\
\ce{VOF3}      & 527.1\cite{doi:10.1021/ic50153a044} & 10 & +5  & 6.8 & -1.3 & -1.2 & 7.1 & -0.9 & -0.9 \\
\ce{VF5}       & 528.9\cite{doi:10.1021/ic50153a044} & 10 & +4  & 7.9 & -1.4 & -1.1 & 8.3 & -1.0 & -0.8 \\
\ce{Ni(C5H5)2} & 859.9\cite{doi:10.1021/j150625a017} & 20 & +2  & 0.8 & -2.2 & -3.6 & 1.5 & -1.6 & -2.9 \\
\hline \hline
\end{tabular}
\caption{Transition Metal (TM) L$_3$ ionization energies for non 18-electron complexes computed with {NO-QDPT}. The electron-count of the given TM complexes as well as the oxidation state (OS) of the TM centers are given. All methods were computed using decontracted aug-cc-p$\omega$CVTZ\cite{balabanov2005systematically} on TM centers and decontracted aug-cc-pVDZ basis set on others. Full results with absolute L$_\text{2,3}$ IPs with SCAN0 data are available in Supporting Information, Table S4-VII.}
\label{tab:tml3edge_non18e}
\end{table}

We only have data for five non-18-electron complexes (Table \ref{tab:tml3edge_non18e}) so it is hard to draw general conclusions. However in the 4 electron deficient (formally d$^0$) complexes, use of HF orbitals causes very large 7-8 eV overestimates, which are greatly reduced by use of DFT.
There is also very little data with which to assess the performance of {NO-QDPT}/DFT for 2p ionization DS (Table \ref{tab:tml3edge_ds}; two \ce{Fe} DS were taken from the lowest dipole-allowed transitions of the gas phase L-edge \ce{Fe(CO)5} and \ce{Fe(C5H5)2} XAS spectra). The DS results appear similar to those discussed previously for 4th row main group elements. The best accuracy is achieved via NO-QDPT/SCAN/DCB-SNSO, with slight error increases across the 1st TM series, as was also seen for L$_\text{3}$ edge prediction.

\renewcommand{\arraystretch}{1.25}
\begin{table}[!t]
\footnotesize
\begin{tabular}{l|r||R{0.4in}|R{0.4in}|R{0.4in}|R{0.4in}|R{0.4in}|R{0.4in}}
\hline \hline
\multicolumn{2}{c||}{}
 & \multicolumn{1}{c|}{HF} 
 & \multicolumn{1}{c|}{SCAN} 
 & \multicolumn{1}{c|}{B3LYP} 
 & \multicolumn{1}{c|}{HF} 
 & \multicolumn{1}{c|}{SCAN}
 & \multicolumn{1}{c}{B3LYP} \\
 \hline
 \multicolumn{2}{c||}{\textbf{Expt. L-edge DS (eV)}} & \multicolumn{3}{c|}{\textbf{X2C Error (eV)}} & \multicolumn{3}{c}{\textbf{DCB-SNSO Error (eV)}} \\
\hline \hline
\ce{Ti(NO3)4}  & 5.9\cite{FORD198375}  & 0.4& 0.5& 0.5& -0.4& -0.4& -0.4\\
\ce{Cr(CO)6}   & 8.8\cite{10.1063/1.462284}  & 0.6& 0.7& 0.6& -0.6& -0.5& -0.5\\
\ce{Fe(CO)5}   & 12.3\cite{godehusen2017iron}$^*$ & 1.0 & 1.2& 1.1& -0.5& -0.4& -0.4\\
\ce{Fe(C5H5)2} & 12.4\cite{godehusen2017iron}$^*$ & 1.0 & 1.1& 1.1& -0.5& -0.4& -0.4\\
\ce{Ni(acac)2} & 17.4\cite{perera1980x}  & 1.3& 1.5& 1.4& -0.7& -0.5& -0.6\\
\hline \hline
\end{tabular}
\caption{
Transition Metal (TM) DS results computed with {NO-QDPT}. All methods were computed using decontracted aug-cc-p$\omega$CVTZ\cite{balabanov2005systematically} on TM centers and decontracted aug-cc-pVDZ basis set on others. The acetylacetonate ligand is abbreviated as acac. $^*$ Fe L-edge DS values are taken from the 1\textsuperscript{st} dipole allowed \emph{excitations} of Ref \citenum{godehusen2017iron}, as results from CEBEs were not available.
}
\label{tab:tml3edge_ds}
\end{table}

\renewcommand{\arraystretch}{1.25}
\begin{table}[t!]
\footnotesize
\begin{tabular}{l|l|r|r|r|r|r|r|r|r}
\hline \hline
 \multicolumn{2}{c|}{} & \multicolumn{1}{c|}{Kr} & \multicolumn{1}{c|}{HBr} & \multicolumn{1}{c|}{\ce{CH3Br}} & \multicolumn{1}{c|}{Kr} & \multicolumn{1}{c|}{Cd} & \multicolumn{1}{c|}{\ce{CH3I}} & \multicolumn{1}{c|}{\ce{CH3I}} & \multicolumn{1}{c}{Xe} \\
 \multicolumn{2}{c|}{} & \multicolumn{1}{c|}{M$_3$} & \multicolumn{1}{c|}{M$_5$} & \multicolumn{1}{c|}{M$_5$} & \multicolumn{1}{c|}{M$_5$} & \multicolumn{1}{c|}{M$_5$} & \multicolumn{1}{c|}{M$_5$} & \multicolumn{1}{c|}{N$_5$} & \multicolumn{1}{c}{N$_5$} \\
 \hline
\multicolumn{2}{c|}{\textbf{Expt. M/N IP (eV)}}   
 & 214.4\cite{PhysRevA.101.042505}
 & 77.3\cite{JJohnson_1997}
 & 76.2\cite{JJohnson_1997}
 & 93.8\cite{king1977investigation}
 & 412.9\cite{banna1978free_zncd}
 & 626.8\cite{forbes2018photoionization}
 & 56.6\cite{forbes2020photoionization} 
 & 67.5\cite{king1977investigation} \\
\hline \hline 
\textbf{Method}  &  \multicolumn{1}{c|}{\textbf{Model} (${\hat{H}}$)} & \multicolumn{8}{c}{Error (eV)}  \\
\hline 
HF    & X2C      & 2.5& -2.0& -2.1& -1.9& -1.3& -1.0& -1.4& -1.3\\
SCAN  & X2C      & -6.2& -1.1& -1.1& -1.1& -2.7& -3.9& -0.6& -0.6\\
SCAN0 & X2C      & -3.8& -1.0& -1.0& -1.0& -2.0& -2.8& -0.5& -0.5\\
B3LYP & X2C      & -6.7& -0.8& -0.8& -0.9& -3.1& -4.6& -0.7& -0.8\\
\hline 
HF    & DCB-SNSO & 2.8& -1.8& -1.9& -1.7& -0.4& 0.3& -1.2& -1.1\\
SCAN  & DCB-SNSO & -6.0& -0.9& -0.9& -0.9& -1.8& -2.6& -0.5& -0.4\\
SCAN0 & DCB-SNSO & -3.6& -0.8& -0.8& -0.8& -1.2& -1.5& -0.3& -0.3\\
B3LYP & DCB-SNSO & -6.5& -0.6& -0.6& -0.7& -2.2& -3.3& -0.6& -0.6\\
\hline \hline
\multicolumn{10}{c}{} \\
\hline \hline
 \multicolumn{2}{c|}{} & \multicolumn{1}{c|}{Kr} & \multicolumn{1}{c|}{HBr} & \multicolumn{1}{c|}{\ce{CH3Br}} & \multicolumn{1}{c|}{Kr} & \multicolumn{1}{c|}{Cd} & \multicolumn{1}{c|}{\ce{CH3I}} & \multicolumn{1}{c|}{\ce{CH3I}} & \multicolumn{1}{c}{Xe} \\
 \multicolumn{2}{c|}{} & \multicolumn{1}{c|}{M$_{2,3}$} & \multicolumn{1}{c|}{M$_{4,5}$} & \multicolumn{1}{c|}{M$_{4,5}$} & \multicolumn{1}{c|}{M$_{4,5}$} & \multicolumn{1}{c|}{M$_{4,5}$} & \multicolumn{1}{c|}{M$_{4,5}$} & \multicolumn{1}{c|}{N$_{4,5}$} & \multicolumn{1}{c}{N$_{4,5}$} \\
 \hline
\multicolumn{2}{c|}{\textbf{Expt. SOC Split (eV)}}   
 & 7.8\cite{PhysRevA.101.042505}
 & 1.0\cite{JJohnson_1997}
 & 1.0\cite{JJohnson_1997}
 & 1.3\cite{king1977investigation}
 & 6.8\cite{banna1978free_zncd}
 & 11.5\cite{forbes2018photoionization}
 & 1.7\cite{forbes2020photoionization} 
 & 2.0\cite{king1977investigation} \\
\hline \hline 
\textbf{Method}  &  \multicolumn{1}{c|}{\textbf{Model} (${\hat{H}}$)} & \multicolumn{8}{c}{Error (eV)}  \\
\hline 
HF    & X2C& 0.5  & 0.5 & 0.5 & 0.6 & 2.8 & 3.0  & 0.4  & 0.4  \\
SCAN  & X2C & 0.6  & 0.5 & 0.5 & 0.6 & 2.9 & 3.1  & 0.5  & 0.5  \\
SCAN0 & X2C & 0.5  & 0.5 & 0.5 & 0.6 & 2.9 & 3.0  & 0.4  & 0.5  \\
B3LYP & X2C & 0.6  & 0.5 & 0.5 & 0.6 & 2.9 & 3.1  & 0.5  & 0.5  \\
HF    & DCB-SNSO & -0.2 & 0.0 & 0.0 & 0.0 & 0.7 & -0.3 & -0.1 & -0.1 \\
SCAN  & DCB-SNSO & -0.1 & 0.0 & 0.0 & 0.0 & 0.7 & -0.2 & 0.0  & 0.0  \\
SCAN0 & DCB-SNSO & -0.1 & 0.0 & 0.0 & 0.0 & 0.7 & -0.2 & 0.0  & -0.1 \\
B3LYP & DCB-SNSO & -0.1 & 0.0 & 0.0 & 0.0 & 0.7 & -0.2 & 0.0  & 0.0 \\
\hline \hline
\end{tabular}
\caption{SOC splittings of M/N edges and corresponding absolute IP for selected gas phase molecules containing beyond 4\textsuperscript{th} row elements, computed through {NO-QDPT}. All methods were computed using decontracted aug-cc-p$\omega$CVTZ-DK\cite{peterson2005systematically, deyonker2007systematically, bross2013correlation} for Kr, Br, I, Xe and Cd, and decontracted cc-pVDZ for else. Full results with absolute IPs are available in Supporting Information, Table S4-VIII.}
\label{tab:mandnedges}
\end{table}

\renewcommand{\arraystretch}{1.25}
\begin{table}[!htb]
\footnotesize
\begin{tabular}{lccc|lccc|lccc}
\hline\hline
\multicolumn{4}{c|}{\ce{CH3I}\cite{forbes2020photoionization} N$_{4,5}$} & \multicolumn{4}{c|}{HBr\cite{JJohnson_1997} M$_{4,5}$} & \multicolumn{4}{c}{HCl\cite{kivilompolo2000cl} L$_{2,3}$} \\
\hline
\textbf{State} & \textbf{Expt.} & \textbf{HF} & \textbf{SCAN} & \textbf{State} & \textbf{Expt.} & \textbf{HF} & \textbf{SCAN} & \textbf{State} & \textbf{Expt.} & \textbf{HF} & \textbf{SCAN}\\
\hline
E$_{\frac{1}{2}}$     & 0.00 & 0.00 & 0.00 & $\Delta_{\frac{5}{2}}$     & 0.00 & 0.00 & 0.00 & $\Pi_{\frac{3}{2}}$      & 0.00 & 0.00 & 0.00 \\
E$_{\frac{3}{2}}$          & 0.19 & 0.18 & 0.17 & $\Pi_{\frac{3}{2}}$          & 0.18 & 0.18 & 0.18 & $\Sigma_{\frac{1}{2}}$ & 0.09 & 0.03 & 0.04 \\
E$_{\frac{1}{2}}$       & 0.35 & 0.29 & 0.29 & $\Sigma_{\frac{1}{2}}$       & 0.30 & 0.31 & 0.30 & $\Pi_{\frac{1}{2}}$      & 1.66 & 1.61 & 1.64 \\
E$_{\frac{3}{2}}$       & 1.73 & 1.69 & 1.73 & $\Delta_{\frac{3}{2}}$       & 1.08 & 1.06 & 1.07 &  &  &  &  \\
E$_{\frac{1}{2}}$          & 1.93 & 1.91 & 1.95 & $\Pi_{\frac{1}{2}}$          & 1.30 & 1.29 & 1.29 &  &  &  &  \\
\hline\hline
$\Gamma$ & $\sim$ 0.2& & & $\Gamma$ & $\sim$ 0.2& & & $\Gamma$ & $\sim 0.1$ & &\\
\hline\hline

\end{tabular}
\caption{Energies (in eV, relative to the lowest energy feature) of the different core-hole states arising from molecular field induced loss of spherical symmetry for the N$_{4,5}$ edge of \ce{CH3I}, M$_{4,5}$ edge of HBr and L$_{2,3}$ edge of HCl. The experimental core-hole state lifetime widths $\Gamma$ are also reported. All calculations utilize NO-QDPT with the DCB-SNSO model. Full results with absolute IPs are available in Supporting Information.}
\label{tab:molecular_field}
\end{table}

\subsection{M- and N- edges of 4\textsuperscript{th} Row and Beyond \label{sec:beyond4th}}

We next explore the performance of NO-QDPT/DFT in predicting CEBEs of d-electrons where SOC couples 5 different spatial orbitals, leading to d$_{5/2}$ and d$_{3/2}$ sublevels. Table~\ref{tab:mandnedges} tabulates M/N-edge SOC splittings and the smaller CEBE of the multiplet (\textit{i.e.,} M$_\text{5}$ of M$_\text{4,5}$, N$_\text{5}$ of N$_\text{4,5}$). The DS results again show improved performance of DCB-SNSO in NO-QDPT over X2C, as well as insensitivity to choice of functional for orbital optimization. 
The CEBEs, however, are systematically underestimated in virtually all cases, similar to behavior observed for the first row TM ligand complexes (Table~\ref{tab:tml3edge_18e}). 
DFT mostly improves over HF but is worse for Kr M$_3$, Cd M$_5$ and \ce{CH3I} M$_5$. This again demonstrates that alternative approaches towards modeling SOC is needed for these heavier elements.
rDCB-SNSO yields very similar results to universal DCB-SNSO for the splittings and the absolute edges (Supporting Information, Table S4-VIII). Finally, we note that the NO-QDPT/DFT results for the Kr M\textsubscript{3} edge are particularly poor. 

\subsection{Molecular field splitting in \ce{CH3I}, \ce{HBr} and \ce{HCl}}
The lack of spherical symmetry in molecular environments formally breaks the degeneracy within a given SOC split multiplet like the 2p$_{3/2}$ orbitals. The resulting molecular field splitting is often smaller than or comparable to the lifetime width of the XPS spectral feature, and as a result is difficult to experimentally to observe. Throughout this work, we have therefore averaged over all possible eigenvalues within a multiplet (such as all four Cl 2p$_{3/2}$ hole states in HCl) to compute DS splittings or CEBEs. For a few small molecules like HBr\cite{JJohnson_1997} and HCl\cite{kivilompolo2000cl} however, it is possible to fit individual ionization contributions to an observed multiplet feature (with tolerable error bars $\le 0.1$ eV). As previously noted, a limitation of NO-QDPT is that it cannot account for dynamic electron correlation contributions to the molecular field induced splitting. These molecules therefore present an opportunity to test the practical implications of neglecting this effect on the quality of predicted XPS spectra. 

The relative energies of the molecular field induced N$_{4,5}$ edge of \ce{CH3I}, M$_{4,5}$ edge of HBr and L$_{2,3}$ edge of HCl are reported in Table \ref{tab:molecular_field}. The agreement is excellent for \ce{CH3I} and HBr, with similarly good performance obtained from \ce{CH3Br} (shown in Supporting Information, Table S4-IX).  However the molecular field splitting is significantly underestimated for HCl, indicating dynamic correlation effects matter more for this system. This means little in practical terms on account of the $\sim 0.1$ eV lifetime broadening of the associated spectral features, which makes them difficult to resolve within the L$_3$ peak, but nonetheless indicates a potential need for caution in modeling very small molecular field induced splittings. This behavior may be analogous to stronger correlation effects in transition metal complexes with smaller ligand field splittings\cite{Atanasov2012,shee2021revealing}. On the other hand, it is encouraging that the larger molecular field splittings in the XUV range (Br M$_{4,5}$ edge and I N$_{4,5}$ edge) are reproduced well by NO-QDPT, as these are more likely to be experimentally resolvable.

\section{Conclusions}\label{sec:conclusion}

In this work, we have proposed, implemented, and benchmarked a new protocol to extend orbital optimized DFT (OO-DFT) to enable quite accurate yet tractable calculations of molecular core-electron binding energies (CEBEs) and doublet splittings (DS) associated with non-zero orbital angular momentum, when important spin-orbit coupling (SOC) effects are at play. This non-orthogonal quasi-degenerate perturbation theory (NO-QDPT) uses OO-DFT to optimize the orbitals of all determinants needed to span the space of possible holes (e.g. 3 sets of spatial orbitals span the 6-dimensional Hilbert space of a 2p hole). Those determinants are used in non-orthogonal configuration interaction (NOCI) to determine the split sublevels, using a relativistic 1eX2C Hamiltonian which includes the SOC effects with screening. The diagonal matrix elements are shifted using the average of the OO-DFT calculations. 

Our main results and conclusions are as follows:
\begin{enumerate}
    \item The SNSO family of SOC Hamiltonians in combination with NO-QDPT/SCAN yielded highly accurate L$_\text{2,3}$ ionization energies and DS for 3\textsuperscript{rd} row main group elements. Absolute CEBEs show $<0.2$ eV error, while the DS values have even smaller ($\sim 0.04$ eV) errors.
    \item {NO-QDPT}/SCAN demonstrates semi-quantitative accuracy on L$_\text{2,3}$ CEBEs and DS of 4\textsuperscript{th} row s-block elements, with errors gradually increasing with atomic number. Results for 3d transition metal complexes show a similar trend.
    \item There are several potential sources of error in the NO-QDPT/DFT approach, including density functional errors, the quality of the core-hole state subspace, errors arising from the combination of NOCI and DFT, and the treatment of the scalar and vector relativistic Hamiltonian terms. Further work is needed to unravel their relative contributions to the largest errors seen for L$_\text{2,3}$ CEBEs and DS of transition metal atoms and heavier elements.
\end{enumerate}

Our promising 2p core ionization results for the lighter elements motivate future work on extending NO-QDPT to L-edge excitations. We intend to approach this challenge by extending 1-center NOCIS\cite{oosterbaan2020generalized}. We are also interested in improved approaches for generating the core-hole orbitals. 
An important question is better understanding the origin of the poorer performance of NO-QDPT/DFT for L-edge CEBEs. Would a better relativistic Hamiltonian improve these results or are other missing effects at play?

\section*{Acknowledgment} 
This work was supported by the Liquid Sunlight Alliance, which is funded by the U.S. Department of Energy, Office of Science, Office of Basic Energy Sciences, Fuels from Sunlight Hub under Award Number DE-SC0021266. Additional funding was provided through the Atomic, Molecular and Optical Sciences (AMOS) program at Lawrence Berkeley Laboratory by the Director, Office of Science, Office of Basic Energy Sciences, of the U.S. Department of Energy, under Contract No. DE-AC02-05CH11231. D.H. was a Stanford Science Fellow for the first stage of this work. The Flatiron Institute is a division of the Simons Foundation.

\section*{Supporting Information}
\begin{itemize}
    \item PDF: Contains supplementary results and discussion on the following --- Theoretical formulation of 1eX2C-Z$_\text{eff}$ Hamiltonian, Computational aspects of NO-QDPT, Alkali atom D-line excitations via NO-QDPT, Experimental references used for main group 3$^\text{rd}$ row elements, Performance of 1eX2C-Z$_\text{eff}$ Hamiltonian for elements beyond third row, and assessment of delocalization error for Kr inner-shell ionization.
    \item XLSX: Basis set convergence test, expanded performance comparison test (including different methods and effective Hamiltonians) for 18-electron series, and detailed results including absolute edges and SOC splitting predictions for all the data presented in the main text. Data in this spreadsheet are noted as Table S4.
    \item ZIP: Geometries of all species considered in XYZ format.
\end{itemize}

\section*{Conflicts of Interest}
M.H.-G. is a part-owner of Q-Chem, which is the software platform in which the developments described here were implemented.

\bibliography{references}
\end{document}


\maketitle
\newpage 

\section{The 1eX2C-Z$_\text{eff}$ Hamiltonian}

The empirical scaling relationship utilized by the 1eX2C-Z$_\text{eff}$ Hamiltonian is described in Refs \citenum{koseki1992mcscf, koseki1995, koseki1998}. These relationships were determined via forming an effective one-electron Breit-Pauli Hamiltonian with an empirical nuclear charge $Z_\text{eff}$ as a fitting parameter:
\begin{align}
    \hat{h}_{1}^\text{Breit-Pauli} \simeq \frac{\alpha^2}{2} \sum_{i,A} \frac{Z_{\rm eff, A}\left( Z \right)}{r_{iA}^3} \hat{L}_{iA} \cdot \hat{S}_{i}
\end{align} where $\alpha$ is the fine structure constant, index $i$ runs over electrons, index $A$ runs over nuclei, $\hat{L}$ represents the orbital angular momentum operator, and $\hat{S}$ represents the spin angular momentum operator. The effective nuclear charges were determined via low-lying excited states of small molecules and their corresponding fine structure splittings. Specifically, MCSCF calculations with empirical $Z_{\rm eff, A}$ were performed to replicate experimental values for low-lying triplet/doublet states, in combination with a non-relativistic Hamiltonian.

To compute the corresponding one-body X2C operator $\hat{h}_1^{\rm X2C-Z_{eff}}$ from the empirical effective nuclear charges, we first compute $\hat{W}_{\rm SO}^{\rm Z_{eff}}$ via
\begin{align}
    \hat{W}_{\rm SO}^{\rm Z_{eff}} &\equiv \hat{\vec{p}} \times \hat{V}^{\rm Z_{eff}} \hat{\vec{p}} \\ 
    \hat{V}^{\rm Z_{eff}} &\equiv \sum_{i,A} \frac{Z_{\rm eff, A}}{r_{iA}}
\end{align} where effective nuclear charge is included in the nuclear attraction operator. This $\hat{W}_{\rm SO}^{\rm Z_{eff}}$ and $\hat{W}_{\rm SF}$ are used to compute $\hat{W}^{\rm Z_{eff}} = \hat{W}_{\rm SF} + i \hat{\vec{\sigma}} \cdot \hat{W}_{\rm SO}^{\rm Z_{eff}}$. Note that the effective nuclear charges do not enter the spin-free part of $\hat{W}^{\rm Z_{eff}}$. This is motivated by the observation that most treatment of effective SOC under 1eX2C formalism attempts to parametrize the spin-dependent part only. Once the matrix $\mathbf{W}^{\rm Z_{eff}}$ is computed, we solve the four-component one-body Dirac equation via:
\begin{align}
    \begin{bmatrix}
        \mathbf{V} & \mathbf{T} \\
        \mathbf{T} & \dfrac{\mathbf{W}^{\rm Z_{eff}}}{4c^2} -\mathbf{T} \\
    \end{bmatrix}
    \begin{bmatrix}
        \mathbf{C}_L \\
        \mathbf{C}_S \\
    \end{bmatrix} &= \epsilon
    \begin{bmatrix}
        \mathbf{S} & 0 \\
        0 & \dfrac{1}{2c^2}\mathbf{T} \\
    \end{bmatrix}
    \begin{bmatrix}
        \mathbf{C}_L \\
        \mathbf{C}_S \\
    \end{bmatrix} 
\end{align}
This yields the corresponding decoupling and renormalization matrices $\mathbf{X}^{\rm Z_\text{eff}}$ and $\mathbf{R}^{\rm Z_\text{eff}}$. These matrices can be used to obtain the desired one-body model X2C operator with effective nuclear charges, $\hat{h}_1^{\rm X2C-Z_{eff}} = \hat{T}^{\rm X2C-Z_{eff}} + \hat{V}^{\rm X2C-Z_{eff}}$, using:
\begin{align}
    \mathbf{T}^{\rm X2C-Z_{eff}} &= \left( \mathbf{R}^{\rm Z_{eff}} \right)^\dagger \left[ \mathbf{X}^{\rm Z_{eff}} + \left( \mathbf{X}^{\rm Z_{eff}} \right)^\dagger \mathbf{T} - \left( \mathbf{X}^{\rm Z_{eff}} \right)^\dagger \mathbf{T} \mathbf{X}^{\rm Z_{eff}} \right] \mathbf{R}^{\rm Z_{eff}} \\
    \mathbf{V}^{\rm X2C-Z_{eff}} &= \left( \mathbf{R}^{\rm Z_{eff}} \right)^\dagger \left[ \mathbf{V} + \frac{1}{4c^2}\left( \mathbf{X}^{\rm Z_{eff}} \right)^\dagger \mathbf{W}^{\rm Z_{eff}} \mathbf{X}^{\rm Z_{eff}} \right] \mathbf{R}^{\rm Z_{eff}} .
\end{align}
The resulting one-electron operator is then combined with non-relativistic two-electron operator as described in main text.

~\newpage
\section{Computational Aspects of {NO-QDPT}}

\subsection{Treatment of the core-hole ROHF MO coefficients}

To properly treat SOC effects within NOCI using $\hat{H}^\textrm{1eX2C-model}$, the core-ionized states $\{\ket{\Psi_{n}}\}$ must be prepared in matrix form suitable for complex Generalized Hartree Fock (cGHF) calculations. For {NO-QDPT}, however, the SOC effects are only introduced while diagonalizing $\hat{H}^\textrm{1eX2C-model}$ but not while optimizing the core-ion states, which are obtained through restricted open-shell HF/DFT calculations. Therefore, the coefficient matrix $\textbf{C}$ for cGHF molecular orbitals is a $\mathbb{C}^2$ matrix of dimensions $2 N_{AO} \times 2 N_{AO}$, whereas the core-ion MO coefficients $\ket{\Psi_{N}}$ are obtained as $\mathbb{R}^2$ matrices of dimensions $N_{AO} \times N_{AO}$ ($N_{AO}$: number of AO basis). This requires a proper transformation of the ROHF MO coefficient matrix ($\mathbf{C}_n \in \mathbb{R}^2$) for it to be compatible with the computation of matrix elements of the 1eX2C-model Hamiltonian matrix ($\mathbf{H}^\textrm{1eX2C-model}$), which is specified below.

\begin{enumerate}
   \item Obtain $n$ states with localized holes $ \{ \ket{\Psi_{n}} \} $, using ROHF (or RO-DFT) calculations with spin-free 1eX2C Hamiltonian $\hat{H}^\textrm{SF-1eX2C}$. The occupied block of the MO coefficients $\mathbf{C}_{\textrm{occ}, n}$ will have dimensions of $N_{AO} \times \left(N_\textrm{occ}-1\right)$, where $N_\textrm{occ}$ is the number of occupied spatial MOs for the ground state RHF calculation (and thus half the number of electrons in ground state).
   
   \item Define hole-reattached determinants $\ket{\Psi^{n}} = \hat{a}_{n}^\dagger \ket{\Psi_{n}}$ and its corresponding occupied MO coefficient matrix as $\mathbf{C}_\textrm{occ}^n$. This is simply done by adding an extra column corresponding to the localized hole on ${C}_n$ and is therefore a $N_{AO} \times N_{occ}$ matrix.

   \item In a cGHF calculation, determinants with $\alpha$ and $\beta$ holes must be represented with different columns of MO coefficient matrices. We obtain the real parts of each MO coefficient for {NO-QDPT} using direct sum:
   \begin{align}
       \mathbf{C}_{2n-1} &= \mathbf{C}_{\textrm{occ}, n} \oplus \mathbf{C}_\textrm{occ}^n \\
       \mathbf{C}_{2n} &= \mathbf{C}_\textrm{occ}^n \oplus \mathbf{C}_{\textrm{occ}, n} .
   \end{align} Both $\mathbf{C}_{2n-1}$ and $\mathbf{C}_{2n}$ have dimensions of $2N_{AO} \times 2N_\textrm{occ}-1$. Note that $\mathbf{C}_{2n-1}$ and $\mathbf{C}_{2n}$ are permutations of each other. For the imaginary part, simply use a zero matrix of dimensions $2N_{AO} \times \left(2N_\textrm{occ}-1\right)$. Use $\mathbf{C}_{2n-1}$ and $\mathbf{C}_{2n}$ in combination with the zero matrix to obtain complex-valued matrices $\mathbf{\tilde{C}}_{2n-1}, \mathbf{\tilde{C}}_{2n} \in \mathbb{C}^2$ with dimensions $2N_{AO} \times 2N_\textrm{occ}-1$.

   \item Our states are now suitable for computing NOCI matrix elements\cite{thom2009hartree} $\mathbf{H}_{AB}^\textrm{1eX2C-model}$ and $\mathbf{S}_{AB}$ using the complex-valued matrices $\mathbf{\tilde{C}}_A$ and $\mathbf{\tilde{C}}_B$ for the determinants $A$ and $B$.
\end{enumerate}

\subsection{Preparation of the ground state}

To obtain excitation or ionization energy, one can either subtract these energies from the true cGHF ground-state (GS) solution or the RHF ground-state solution with a perturbative correction of SOC at the ground-state. In this work, we chose the latter as it permits the use of Kohn-Sham determinants without implementing a non-collinear form of the DFT kernels. Same treatment as previous section can be done for compatibility with cGHF calculation but in a much simpler form, which would involve a matrix tensor product between the $2 \times 2$ identity matrix and the ground state RHF MO coefficient matrix.

\subsection{Preparation of the localized hole determinants}

For NO-QDPT, each core-hole must be localized and the resulting determinants be orthogonal (or near-orthogonal) to each other. We found the following strategies to be useful in achieving these conditions:

\begin{enumerate}
    \item Feeding in a ground state guess obtained with a small electric field. This has an effect of breaking the (near-) degeneracy of inner-shell orbitals and orienting them in specific directions. This enables us to obtain the states of interest under the spin-free Hamiltonian without mixing of inner-shell orbitals with different $L_z$ values.
    \item For systems with multiple ionization/excitation centers (\textit{e.g.,} \ce{Cl2}): Performing a Boys Localization on a converged ground state calculation (after SCF) and then feeding in this guess. Alternatively, applying a small electric field can also break the spatial symmetry of the orbitals as well. 
    \item For Kohn-Sham determinants: Computing exchange-correlation contributions with finer grids, to minimize errors arising from the lack of rotational invariance.
\end{enumerate}

To examine the orthogonality of the Slater determinants within the NOCI subspace, we performed singular value decomposition (SVD) on the NOCI overlap matrices ($\mathbf{S}$) for the 3rd row species noted in Fig. 2 of main text and found that all the singular values are within $[0.99, 1.01]$ (Table S4-V). This indicates that there are virtually no linear dependencies within the NOCI subspace. 
However, as this check must be done \textit{a posteriori} after optimization of all core-hole states, a constrained optimization scheme where strict orthogonality between all core-hole states involved in a NOCI calculation are imposed throughout the $\Delta$SCF iterations would be helpful and desirable.

~\newpage
\section{Computational Details for $\Delta$cGHF calculation}

To compute the L$_2$ and L$_3$ ionization energies via state-specific $\Delta$SCF calculations in complex-generalized Hartree Fock ($\Delta$cGHF), six 2p-like spin-orbitals of a ground state calculation were ionized and optimized. Since $\Delta$cGHF calculations were performed for \ce{Ar, HCl, H2S, PH3,} and  \ce{SiH4}, the ionization targets were the 5$^\text{th}$ to 10$^\text{th}$ lowest energy orbitals. 
The ionized states were optimized via the Maximum Overlap Method.\cite{gilbert2008self}

The optimized $\Delta$cGHF calculations led to six different states, with four states of near-identical energies corresponding to L$_3$ ionization and two higher energy states of near-identical energies corresponding to L$_2$ ionization. We note that the lack of perfect fourfold degeneracy within the L$_3$ band and perfect twofold degeneracy within the L$_2$ band arises from spherical symmetry breaking due to molecular field (and hence does not occur for Ar). The band average ionization energies were reported as $\Delta$cGHF L$_2$ and L$_3$ CEBEs. 
Finally, the difference of L$_2$ IP and L$_3$ IP was reported as the $\Delta$cGHF DS value.

~\newpage
\section{Alkali Atom D-lines Splitting}

\renewcommand{\arraystretch}{1.25}
\begin{table}[b!]
\footnotesize
\begin{tabular}{l|l|R{0.8in}|R{0.8in}|R{0.8in}|R{0.8in}|R{0.8in}}
\hline \hline
\multicolumn{2}{c|}{} & \multicolumn{1}{c|}{Li} & \multicolumn{1}{c|}{Na} & \multicolumn{1}{c|}{K} & \multicolumn{1}{c}{Rb} & \multicolumn{1}{c}{Cs} \\
\hline
\multicolumn{2}{c|}{\textbf{Expt. (cm$^\text{-1}$)}} 
 & 0.34\cite{scherf1996re} 
 & 17.20\cite{juncar1981absolute} 
 & 57.71\cite{falke2006transition} 
 & 237.60\cite{ye1996hyperfine, barwood1991frequency}
 & 554.04\cite{udem1999absolute, udem2000absolute} \\
\hline \hline
\textbf{Method}  &  \multicolumn{1}{c|}{\textbf{Model} (${\hat{H}}$)} & \multicolumn{5}{c}{\textbf{Error (\%)}} \\
\hline \hline
HF       & X2C      & 357.0  & -4.7  & -27.7 & -36.5 & -42.3 \\
SCAN     & X2C      & 217.0  & 42.9  & 11.4  & 1.0   & -6.4  \\
SCANh    & X2C      & 166.9  & 39.3  & 9.1   & -0.7  & -7.9  \\
SCAN0    & X2C      & 106.9  & 34.1  & 6.0   & -3.4  & -10.4 \\
B3LYP    & X2C      & 712.5  & 63.2  & 28.9  & 15.3  & 4.5   \\
$\omega$B97X-D3 & X2C      & 631.3  & 67.2  & -8.1  & -18.7 & -26.0 \\
$\omega$B97X-V  & X2C      & 462.1  & 40.9  & 5.3   & -7.3  & -17.1 \\
BHHLYP   & X2C      & 1038.2 & 41.9  & 10.9  & -1.5  & -10.7 \\
\hline
HF       & DCB-SNSO & -95.5  & -30.4 & -39.0 & -41.6 & -45.4 \\
SCAN     & DCB-SNSO & 161.5  & 4.4   & -6.0  & -7.1  & -11.5 \\
SCANh    & DCB-SNSO & 91.6   & 1.7   & -7.9  & -8.7  & -12.8 \\
SCAN0    & DCB-SNSO & 4.5    & -2.1  & -10.5 & -11.2 & -15.2 \\
B3LYP    & DCB-SNSO & 466.6  & 19.2  & 8.8   & 6.0   & -1.2  \\
$\omega$B97X-D3 & DCB-SNSO & 444.8  & 45.3  & -22.5 & -25.2 & -30.0 \\
$\omega$B97X-V  & DCB-SNSO & -33.4  & 2.9   & -11.2 & -14.7 & -21.6 \\
BHHLYP   & DCB-SNSO & 905.4  & 3.6   & -6.4  & -9.4  & -15.5 \\
\hline \hline
\end{tabular}
\caption{Gas phase alkali D-line splittings calculated through NO-QDPT. Orbitals were computed using the decontracted aug-cc-p$\omega$CVTZ-X2C\cite{hill2017gaussian} (\ce{K, Rb, Cs}) basis set and decontracted aug-cc-pVTZ\cite{prascher2011gaussian} (\ce{Li, Na}) basis set. The predicted values and the experiment are in wavenumbers (cm\textsuperscript{-1}) and the errors are reported in percent (\%). Full results are available in our separately provided spreadsheet.}
\label{tab:dlines}
\end{table}

We have also extended our approach to calculating D1-D2 splitting of alkali atoms. This excitation corresponds to the valence $ns^1 \rightarrow np^1$ excitation, where a splitting between two different D lines can be observed due to SOC. Table~\ref{tab:dlines} shows the \% errors in gas phase D-line splittings computed through {NO-QDPT}, vs experiment.
We note that while percentage error reported for D-line splitting themselves look sub-optimal, this is largely due to the small scale of the experimental energies we are computing the \% error against (as small as $0.34$ cm$^{-1}$ for Li, while  typical thermal fluctuation $k_B T \sim 200$ cm\textsuperscript{-1} for 300 K). SCAN0 reports merely a $\sim$15\% or lower error with DCB-SNSO, a remarkable performance considering the very small magnitude of the energy scale we are investigating. 

\section{Full Results for L$_\text{2,3}$ Edges for 3$^\text{rd}$ Row Element}

Table \ref{tab:expt3rdrow} lists the molecules containing 3$^\text{rd}$ row elements selected to evaluate the performance of NO-QDPT for the 3$^\text{rd}$ row elements in this work.
Experimental results for L$_\text{2,3}$ edge IPs as well as its double splitting are given. For detailed computation results, refer to the separate XLSX spreadsheet provided (Table S4-IV).

\renewcommand{\arraystretch}{1.25}
\begin{table}[H]
\begin{tabular}{l|r|r|r|c}
\hline \hline
\textbf{18-electron Series} & \multicolumn{1}{l|}{Expt. L$_2$ IP (eV)} & \multicolumn{1}{l|}{Expt. L$_3$ IP (eV)} & \multicolumn{1}{l|}{Expt. DS (eV)} & \multicolumn{1}{l}{Ref.} \\
\hline \hline
\ce{Ar}   & 250.78 & 248.63 & 2.15 & \citenum{king1977investigation} \\
\ce{HCl}  & 209.03 & 207.40 & 1.63 & \citenum{shaw1984inner} \\
\ce{H2S}  & 171.56 & 170.36 & 1.20 & \citenum{hudson1994high} \\
\ce{PH3}  & 137.95 & 137.05 & 0.90 & \citenum{liu1990high} \\
\ce{SiH4} & 107.80 & 107.20 & 0.60 & \citenum{hayes1972absorption} \\
\hline \hline
\textbf{Others} & \multicolumn{1}{l|}{Expt. L$_2$ IP (eV)} & \multicolumn{1}{l|}{Expt. L$_3$ IP (eV)} & \multicolumn{1}{l|}{Expt. DS (eV)} & \multicolumn{1}{l}{Ref.} \\
\hline \hline
\ce{CH3Cl} & 207.90 & 206.26 & 1.64 & \citenum{aitken1980electron} \\
\ce{ClF} & 210.83 & 209.18 & 1.65 & \citenum{aitken1980electron} \\
\ce{ClF3} & 214.62 & 213.02 & 1.60 & \citenum{sze1989inner} \\
\ce{Cl2} & 209.45 & 207.82 & 1.63 & \citenum{aitken1980electron} \\
\ce{CCl4} & 208.73 & 207.04 & 1.69 & \citenum{aitken1980electron} \\
\ce{SO2} & 176.00 & 174.80 & 1.20 & \citenum{krasnoperova1976fine} \\
\ce{CSO} & 171.80 & 170.60 & 1.20 & \citenum{krasnoperova1977ii} \\
\ce{CS2} & 171.00 & 169.80 & 1.20 & \citenum{krasnoperova1977ii} \\
\ce{SF6} & 181.48 & 180.27 & 1.21 & \citenum{hudson1993high} \\
\ce{\textbf{S}PF3} $^\text{a}$ & 171.06 & 169.86 & 1.20 & \citenum{neville1998inner} \\
\ce{PF3} & 141.97 & 141.04 & 0.93 & \citenum{ishiguro1987high} \\
\ce{P(CH3)3} & 136.85 & 135.95 & 0.90 & \citenum{liu1990high} \\
\ce{P(CF3)3} & 139.35 & 138.45 & 0.90 & \citenum{liu1990high} \\
\ce{PCl3} & 140.50 & 139.60 & 0.90 & \citenum{ishiguro1987high} \\
\ce{PF5} & 145.14 & 144.31 & 0.83 & \citenum{hu2007high} \\
\ce{OPF3} & 143.86 & 142.96 & 0.90 & \citenum{neville1998inner} \\
\ce{Si(CH3)4} & 106.55 & 105.94 & 0.61 & \citenum{bozek1987high} \\
\ce{SiCl4} & 110.76 & 110.17 & 0.59 & \citenum{bozek1987high} \\
\ce{SiF4} & 112.20 & 111.60 & 0.60 & \citenum{friedrich1980overlapping} \\
\ce{Si(OCH3)4} & 108.03 & 107.42 & 0.61 & \citenum{sutherland1993si} \\
\hline \hline
\end{tabular}
\caption{Experimental gas phase L-edge values for 3$^\text{rd}$ row element containing molecules tested in this work. $^\text{a}$The experimental value given for \ce{SPF3} corresponds to ionization of sulfur.}
\label{tab:expt3rdrow}
\end{table}

\begin{figure}[t!]
    \centering
    \includegraphics[width=0.8\linewidth, trim=100 0 100 30]{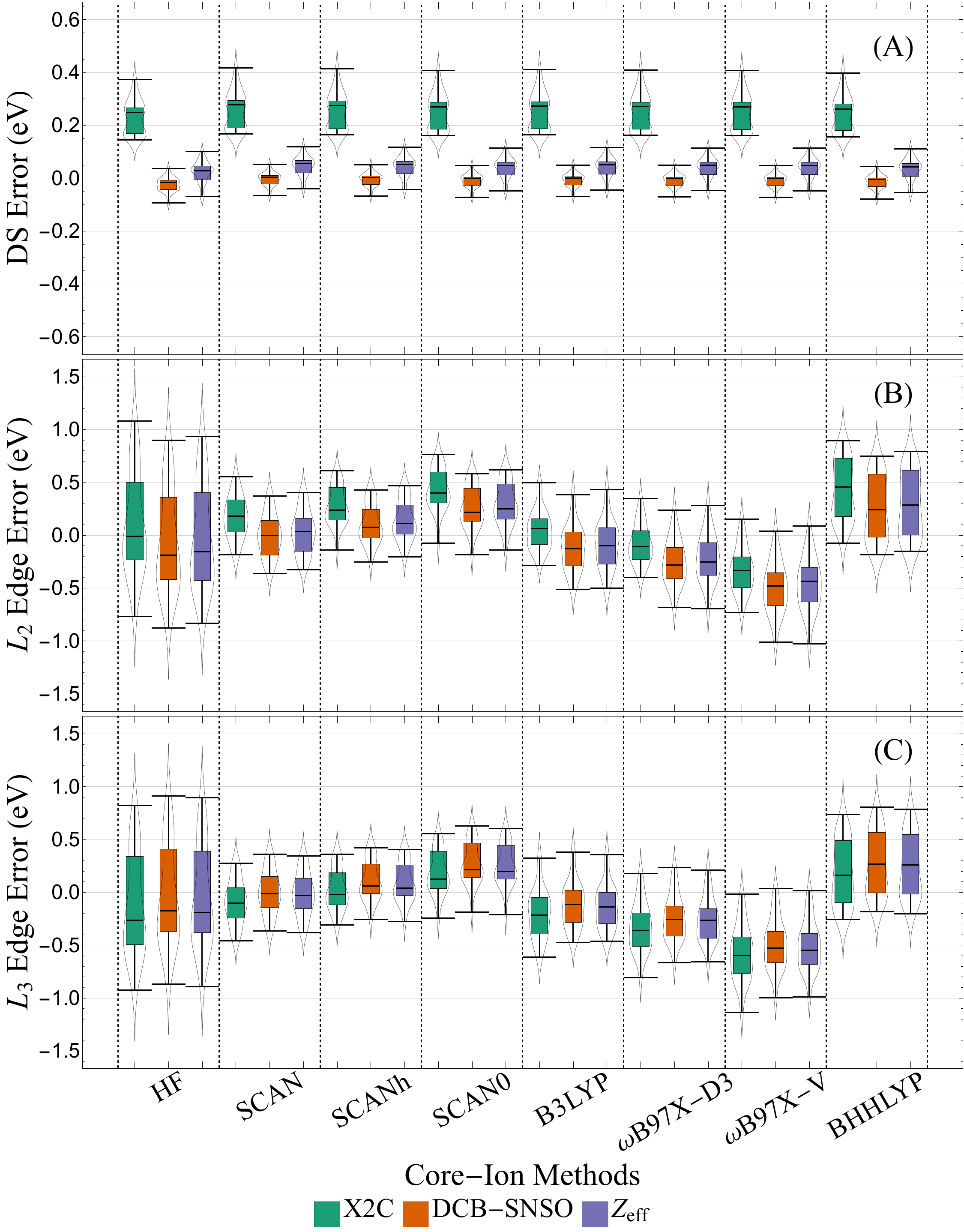}
\caption{Full violin plot results for the gas phase 3\textsuperscript{rd} row element L$_\textrm{2,3}$-edges. Boxes span the 50\% as the interquartile range with the median noted, and the underlying contours show the distribution of the data.}
\label{fig:3rdrow_si}
\end{figure}

\newpage
\section{Performance of Z$_\text{eff}$ Hamiltonian with NO-QDPT for elements beyond 3$^\text{rd}$ Row}

We briefly describe the performance of Z$_\text{eff}$ beyond third row elements in predicting the L$_\text{2,3}$ edge DS. The estimate of L$_\text{2,3}$ edge DS with Z$_\text{eff}$ Hamiltonian is provided in Table \ref{tab:zeffbeyond3rdrow}. Compared against the DCB-SNSO Hamiltonian, the performance of Z$_\text{eff}$ Hamiltonian is notably degraded as the nuclear charge is increased. Especially, the doublet splitting values for the transition metals are erratic, reaching error of $>3$ eV for DS for Kr. Similar error trends are also observed in the absolute L$_\text{3}$ edge values, though with a reduced magnitude. While the results show that early main group parametrization of the effective nuclear charges\cite{koseki1995} is relatively better than the transition metal parametrization\cite{koseki1998}, the lack of consistent performance as well as its worse performance when compared to DCB-SNSO parametrization indicates that effective nuclear charge based approach is less robust.

\renewcommand{\arraystretch}{1.25}
\begin{table}[!b]
\footnotesize
\begin{tabular}{l|R{0.5in}|R{0.5in}|R{0.5in}}
\hline \hline
\multicolumn{2}{c|}{\textbf{Expt. L$_3$ (eV)}} 
   & \multicolumn{1}{c|}{\textbf{Z$_\text{eff}$ Error (eV)}} 
   & \multicolumn{1}{c}{\textbf{DCB-SNSO Error (eV)}} \\
\hline
\ce{Ca}\cite{wernet1998determination}      & 357.6  & -0.6 & -0.4 \\
\ce{Cr(CO)6}\cite{10.1063/1.462284}        & 582.0  &  0.3 &  0.7 \\
\ce{Fe(CO)5}\cite{doi:10.1021/ja00504a017} & 715.8  & -0.3 &  0.2 \\
\ce{Zn}\cite{banna1978free_zncd}           & 1029.1 & -0.6 &  0.2 \\
\ce{GeH4}\cite{venezia1987near}            & 1225.7 & -0.3 &  0.7 \\
\ce{Kr}\cite{dragoun2004increased}         & 1679.2 &  0.3 &  1.8 \\
\hline
\multicolumn{2}{c|}{\textbf{Expt. DS (eV)}} 
   & \multicolumn{1}{c|}{\textbf{Z$_\text{eff}$ Error (eV)}} 
   & \multicolumn{1}{c}{\textbf{DCB-SNSO Error (eV)}} \\
\hline
\ce{Ca}\cite{wernet1998determination}            & 3.4  & 0.7 & 0.1 \\
\ce{Cr(CO)6}\cite{10.1063/1.462284}              & 8.8  & 0.6 & -0.6 \\
\ce{Fe(CO)5}\cite{godehusen2017iron}$^\text{,a}$ & 12.3 & 1.0 & -0.7 \\
\ce{Zn}\cite{banna1978free_zncd}                 & 23.2 & 2.1 & -0.4 \\
\ce{GeH4}\cite{venezia1987near}                  & 31.1 & 2.4 & -0.7 \\
\ce{Kr}\cite{dragoun2004increased}               & 52.7 & 3.2 & -1.4 \\
\hline \hline
\end{tabular}
\caption{L$_3$ edge and L$_\text{2,3}$ DS values for elements beyond 3$^\text{rd}$ row using NO-QDPT and HF core-ions. The performance based on Z$_\text{eff}$ Hamiltonian is compared against that of DCB Hamiltonian.  $^\text{a}$The experimental value of DS given for \ce{Fe(CO)5} corresponds to the DS of the lowest dipole-allowed transition. Same basis sets were used for each molecular species as shown in main text.}
\label{tab:zeffbeyond3rdrow}
\end{table}

The poor performance of the Z$_\text{eff}$ might mostly arise from shortcomings in its parametrization, which is based on valence properties of diatomic hybrides for main-group elements calculated with MCSCF/SBK(d,p).\cite{koseki1992mcscf,koseki1995} Z$_\text{eff}$ approach is therefore agnostic to orbital angular momentum, screening the full SOC operator by the same scaling factor. In contrast, SNSO-based approaches compare orbital energies between four-component and two-component calculations in order to propose a shell-dependent scaling factor. A better parametrization of Z$_\text{eff}$ factors with more sophisticated SOC Hamiltonian with quantities more relevant to core-level splitting may lead to a better performance. As such, we only report results from the SNSO-based parametrization schemes on the main text.

\renewcommand{\arraystretch}{1.25}
\section{Delocalization Errors for Kr Inner-Shell Ionization}
\begin{table}[H]
\centering
\begin{tabular}{lcccc}
\hline \hline
Orbital & a & b & $r^2$ & Non-relativistic binding energy (eV) \\
\hline
1s & 0.012 & 0.989 & 0.9967 & 14095.9 \\
2s & 0.016 & 0.984 & 0.9999 & 1851.4 \\
2p & 0.020 & 0.980 & 0.9999 & 1678.5 \\
3s & 0.043 & 0.957 & 1.0000 & 272.6 \\
3d & 0.111 & 0.889 & 0.9999 & 95.4 \\
4s & 0.186 & 0.813 & 0.9999 & 28.4 \\
4p & 0.312 & 0.688 & 1.0000 & 14.1 \\
\hline \hline
\end{tabular}
\caption{Deviations from linearity\cite{perdew1982density} for fractional ionization of Kr orbitals with nonrelativistic, spin-unrestricted SCAN calculations. Following Ref. \citenum{hait2018delocalization}, the SCAN energies $E(x)$ upon removal of $x$ fractional alpha electrons from the specified orbital are approximated with $E(0)+x(ax+b)(E(1)-E(0))$ (where $E(0)$ is the energy of the neutral atom, and $E(1)$ the energy after the complete removal of an electron). The corresponding $a,b$ are found from linear fits of $\dfrac{1}{x}\dfrac{E(x)-E(0)}{E(1)-E(0)}$ vs $x$, and are reported below along with the corresponding $r^2$ values for the linear fits. The inner-shell orbitals have the smallest curvatures $a$, indicating lower deviation from piecewise linearity. The $3p$ ionization led to several convergence failures, and is thus not reported. }
\label{tab:orbital_data}
\end{table}

\newpage \newpage
\bibliography{references}